\documentclass{aa}

\usepackage{amssymb}
\usepackage{amsmath}
\usepackage{graphicx}
\usepackage{natbib}
\usepackage{multirow}
\usepackage{txfonts}
\usepackage{microtype}
\usepackage{color}
\usepackage[colorlinks=true,citecolor=blue,linkcolor=blue]{hyperref}
\usepackage{epstopdf}
\usepackage[flushleft]{threeparttable}
\usepackage{booktabs}
\usepackage{subfigure}

\begin{document}

\title{Kinematic constraints beyond $z\simeq0$ using calibrated GRB correlations}
\titlerunning{Kinematic constraints beyond $z\simeq0$ using calibrated GRB correlations}
\authorrunning{Luongo, Muccino}

\author{Orlando Luongo\inst{1,2,3}, Marco Muccino\inst{1}}
\institute{Istituto Nazionale di Fisica Nucleare (INFN), Laboratori Nazionali di Frascati, 00044 Frascati, Italy.\\
Scuola di Scienze e Tecnologie, Universit\`a di Camerino, 62032 Camerino, Italy.\\
NNLOT, Al-Farabi Kazakh National University, Al-Farabi av. 71, 050040 Almaty, Kazakhstan.}

\abstract
{The dynamics of the Universe are revised using high-redshift data from gamma-ray bursts to constrain cosmographic parameters by means of model-independent techniques.
}
{
Considering samples from four gamma-ray burst correlations and two hierarchies up to $j_0$ and $s_0$, respectively, we derived limits over the expansion history of the Universe. Since cosmic data span outside $z\simeq0$, we investigated additional cosmographic methods such as auxiliary variables and Pad\'e approximations
}
{
Bezi\'er polynomials were employed to calibrate our correlations and heal the circularity problem. Several Markov chain Monte Carlo simulations were performed on the model-independently calibrated \textit{\emph{Amati}}, \textit{\emph{Ghirlanda}}, \textit{\emph{Yonetoku},} and \textit{\emph{combo}} correlations to obtain $1$--$\sigma$ and $2$--$\sigma$ confidence levels and to test the standard cosmological model.
}
{
Reasonable results are found up to $j_0$ and $s_0$ hierarchies, respectively,  only partially alleviating the tension on local $H_0$ measurements as $j_0$ hierarchy is considered. Discussions on systematic errors have been extensively reported here.
}
{
Our findings show that the $\Lambda$CDM model is not fully confirmed using gamma-ray bursts. Indications against a genuine cosmological constant are summarized and commented on in detail.
}


\offprints{\href{mailto:orlando.luongo@lnf.infn.it}{\nolinkurl{orlando.luongo@lnf.infn.it}},\\
\href{mailto:marco.muccino@lnf.infn.it}{\nolinkurl{marco.muccino@lnf.infn.it}}}

\maketitle

\section{Introduction}

Modern observations increasingly shed light on the properties of the Universe's constituents \citep{Riess1998,Perlmutter1999,Planck2018,2019NatAs...3..891V}. Even though data are more and more accurate, current understanding of cosmology suggests that most of the Universe's content is in the form of dark energy ($\sim68\%$) and dark matter ($\sim23\%$). Unveiling whether dark energy is a cosmological constant\footnote{Standard indicators suggest that  the Universe is currently undergoing an accelerated phase due to an unknown dark energy term  \citep{reviewDE1,reviewDE2}.
} and understanding dark matter's nature\footnote{Dark matter is essential to describing structures and providing enough matter in perturbation theory.} are two of today's most complex challenges \citep{2009JCAP...09..032B,bertone,nostro}.
For decades, disclosing Universe dynamics and clustering properties of its constituents has lead to the search for three quantities: \emph{\emph{the current Hubble rate}, \emph{the present deceleration parameter value, and}  \emph{the amount of visible and invisible matter}} \citep{1998ApJ...492...29M,2000JHEP...06..006B,2000PhRvD..61b3503G,2018PhRvD..97l3521S,2020MNRAS.tmpL..39B}. Although data from standard indicators have progressively exceeded the redshift limit of our local Universe ($z \simeq 0$) measurements are usually jeopardized by postulating the model \emph{\emph{a priori}}. Thus, model-independent treatments that permit us to fix bounds over the dark energy and dark matter contributions have become much more relevant \citep{2000IJMPD...9..373S,PhysRevD.78.103502,2012PhRvD..86l3516A,2012JCAP...06..036S,2014PhRvD..89b3503Y,2014ApJ...793L..40S,2016JCAP...12..042D,2016IJGMM..1330002D}, as they are fascinating topics providing subtle results. The purpose of these prescriptions is to frame out all dynamics, without postulating the cosmological model \emph{\emph{a priori}}. Among several treatments, cosmography aims to fit the observable quantities of interest by means of the Taylor series up to a certain order \citep{2013Galax...1..216C,2014PhRvD..90d3531A}. Cosmographic expansions intrinsically own the idea of model-independently fitting data by fixing limits over scale factor derivatives, expanding around $z\simeq0$. The elegance of cosmography clashes with current surveys that span beyond redshift intervals where cosmography turns out to be predictive \citep{PhysRevD.78.063501}. Due to this restriction over $z,$ the need for a \emph{\emph{high-redshift cosmography}} that involves redshifts outside the local Universe is at an all-time high \citep{2008CQGra..25p5013C}. While a self-consistent approach to high-redshift cosmography is far out of reach, one requires high-redshift standard indicators alongside current approaches to fit data at $z>1$. A possible relevant high-redshift indicator is the presence of gamma-ray bursts (GRBs), which represent the most powerful explosions in the Universe \citep{Meszaros2006}. Their use in cosmology would be a major improvement on currently known outcomes, since they likely
originate from black hole formation, emit huge amounts of energy up to $10^{54}$ erg, and are observable from larger distances than supernovae (SNe). Understanding if GRBs can be suitable variants for standard candles, meaning, if they can be considered as standard indicators, is an open challenge for astrophysics due to the well-known \emph{\emph{circularity problem}}. Moreover, matching universe kinematics with GRB data would give new insights into cosmodynamics. However, relating cosmography to GRBs is a difficult task for two reasons: 1)  GRBs are not yet pure standard candles, and 2) cosmography is accurate for $z\ll1$ only, whereas GRBs are objects that severely exceed the limit $z\simeq0$. To fix these issues, we must a) find a procedure to calibrate GRBs\footnote{To enable them to be quasi-standard candles.}, b) manipulate GRB data points and make them suitable for truncated cosmographic series \citep[see, e.g.,][]{2015NewAR..67....1W}. On the one hand, in order to explore the first issue, one requires a self-consistent GRB theoretical model that is intrinsically capable of connecting all GRB photometric and spectral properties. Despite this caveat being far from clarified, there are consolidate GRB correlations that try to relate photometric and  spectral properties. On the other hand, to enable cosmography to be predictive, two methods are currently available: auxiliary variables and rational approximations of cosmic distances. Both of them permit one to exceed the limit $z\ll1$ for fitting procedures.

Hence, the aim of this work is twofold. First, we considered the mostly used GRB correlations with the lowest data-point spread: \emph{\emph{\emph{\emph{\emph{\emph{\emph{\textit{\emph{Amati}}, \textit{\emph{Ghirlanda}}, \textit{\emph{Yonetoku},}}}} and \textit{\emph{combo}}}}}} correlations.
Then, we overcame the \emph{\emph{circularity problem}} by calibrating the above correlations through a purely model-independent technique at
low redshift based on B\'ezier's approximants and first presented in \citet{2019MNRAS.486L..46A}. Once the correlations were calibrated, our first task was to model-independently enable GRBs to be quasi-standard candles. We thus considered three approaches to fit the Universe's dynamics by means of kinematics. To do so, we employed the first approach purely based on genuine low-redshift cosmography. The other two options resulted from a high-redshift cosmography scenario and consist of using auxiliary variables and optimal rational approximations of the Taylor expansions inferred from Pad\'e polynomials, respectively. We therefore worked with two  hierarchical scenarios, initially involving the sets $\mathcal A_1\equiv \{h_0,q_0,j_0\},$ and $\mathcal A_2\equiv \{h_0,q_0,j_0, s_0\}$ later. This was done by fixing the orders of our expansions. We performed our statistical analyses  using the most recent Pantheon survey of SNeIa, GRB data sets from each of the aforementioned correlations, and the BOSS measurements for Baryonic acoustic oscillations (BAO). We performed a Monte Carlo analysis based on the widely adopted Metropolis-Hastings algorithm, with corresponding contours reported up to $2$--$\sigma$ confidence levels. Although preliminary, our analyses show that the use of GRBs does not fully confirm the hypothesis that dark energy is a pure cosmological constant. The discrepancies are discussed in terms of evolving dark energy or by considering the effects of spatial curvature that have been conventionally neglected in our treatments. In this respect, an in-depth discussion on systematics and a comparison with previous works have been reported, indicating the goodness of our outcomes and the consequences on cosmography.

The paper is structured as follows. In Section 2, we highlight the main features concerning the GRB correlations. In Section 3, we propose the cosmographic method, and we split its use between low- and high-redshift cosmology, respectively. We thus introduce the auxiliary variable and rational approximant methods. In Section 4, we describe the B\'ezier calibration, overcoming the circularity problem previously discussed throughout the text. Afterward, in the same section, we provide details of our experimental analysis, describing each test performed here. In Section 5, we summarize our results, and we propose a theoretical description of our outcomes by means of a slightly evolving scalar field that mimicks dark energy. Finally, in Section 6, we state our conclusions to and perspectives on our approaches.


\section{GRB correlations}

Gamma-ray burst correlations are of utmost importance since they relate observable quantities among them and represent strategies to enable GRBs to be used as distance indicators \citep[for details, please refer to the following works:][]{Amati2002,Ghirlanda04,Amati2008,Schaefer2007,CapozzielloIzzo2008,Dainotti2008,Bernardini2012,AmatiDellaValle2013,Wei2014,Izzo2015,Demianski17a,Demianski17b}. Calibration procedures thus become essential to healing the circularity problem, plaguing each correlation. The circularity issue consists of the fact that
GRB distances are typically obtained through photometry only, and that they depend on the cosmological parameters via $H(z)$, which is the cosmological model that describes Universe dynamics at large scales. So, using GRBs to disclose the Universe's expansion history but calibrating it with a model \emph{\emph{would a priori}}  represent a tricky circularity. This issue is not easy to overcome because there is no low-redshift set of GRBs to achieve a cosmology-independent calibration. The common approach to overcoming circularity is to forcibly add further data at lower redshifts \citep[see, e.g.,][and references therein]{Demianski17a}.

In our work, we employed a new technique that is capable of reconstructing low-redshift correlations in a model-independent way. First, we summarize below the most suitable correlations employed in the literature, such as the ones in which the spread of each GRB point is not as large as it is in other correlations. Later, we describe how to calibrate them by means of B\'ezier polynomials.


\subsection{The \textit{Amati} correlation}

The most investigated correlation in GRB literature is the $E_{\rm p}$--$E_{\rm iso}$ or \emph{\emph{\textit{\emph{Amati}} }}correlation \citep{Amati2002,Amati2008,AmatiDellaValle2013,Demianski17a,Dainotti18}:
\begin{equation}
\label{Amatirel}
\log\left(E_{\rm p}/{\rm keV}\right)= a_0 + a_1 \left[\log\left(E_{\rm iso}/{\rm erg}\right)-52\right]\,,\end{equation}
where $a_0$ and $a_1$ are the two calibration constants.
This correlation involves prompt emission observables: the rest-frame peak energy $E_{\rm p}$ of the $\nu F_\nu$ energy spectrum and the isotropic radiated energy in $\gamma$-rays $E_{\rm iso}=4\pi d_{\rm L} S_{\rm bolo}(1+z)^{-1}$, in which $d_{\rm L}$ is the luminosity distance to the source, $S_{\rm bolo}$ the bolometric fluence computed from the integral of the $\nu F_\nu$ spectrum in the rest-frame $1$--$10^4$~keV energy band, and the factor $(1+z)^{-1}$ transforms the observed GRB duration into the rest-frame one.
The \textit{\emph{Amati}} correlation is characterized by an extra source of variability $\sigma_{\rm a}$, due to the contribution of hidden variables that we cannot observe directly \citep{Dago2005}.


\subsection{The \textit{Ghirlanda} correlation}

Theoretical and observational arguments are in favor of the possible jetted nature of GRBs \citep{2015PhR...561....1K}. In this respect, the GRB prompt emission energy has to be corrected for a collimation factor $f=1-\cos\theta_{\rm jet}$, where $\theta_{\rm jet}$ is the jet opening angle inferred from an achromatic jet break observed in the GRB afterglow light curve at a characteristic time $t_{\rm jet}$ \citep[see, e.g.,][]{Sarietal1999,2001ApJ...562L..55F}.
However, the estimate on $\theta_{\rm jet}$ depends on specific assumptions on the circumburst medium.
We thus chose the case of a homogeneous medium \citep{2001ApJ...562L..55F}. By correcting the radiated energy as $E_\gamma=f\cdot E_{\rm iso}$, we obtain the the $E_{\rm p}$--$E_\gamma,$ or \textit{\emph{"Ghirlanda"}} correlation \citep{Ghirlanda2004,2007A&A...466..127G}:
\begin{equation}
\label{Ghirlandarel}
\log\left(E_{\rm p}/{\rm keV}\right)= b_0 + b_1 \left[\log\left(E_\gamma/{\rm erg}\right)-50\right]\,,
\end{equation}
where $b_0$ and $b_1$ are the two calibration constants.
The correlation features an extra scatter: $\sigma_{\rm b}$.


\subsection{The \textit{Yonetoku} correlation}

The Yonetoku correlation involves $E_{\rm p}$ and the peak luminosity $L_{\rm p}$ being computed from the time interval of $1$~s around the most intense peak of the burst light curve. The peak luminosity is computed as $L_{\rm p}=4\pi d_{\rm L} F_{\rm p}$, where $F_{\rm p}$ is the observed  $1$~s peak flux in the rest frame $30$--$10^4$~keV energy band \citep{Yonetoku2004,2019NatCo..10.1504I}. The $L_{\rm p}$--$E_{\rm p,}$ or \textit{\emph{"Yonetoku"}} correlation is written as
\begin{equation}
\label{Yonetokurel}
\log\left(L_{\rm p}/{\rm erg/s}\right) - 52 = m_0 + m_1 \log\left(E_{\rm p}/{\rm keV}\right)\,,
\end{equation}
where $m_0$ and $m_1$ are the two calibration constants. The \textit{\emph{Yonetoku}} correlation features an extra scatter: $\sigma_{\rm m}$.


\subsection{\textit{Combo} correlation}

The \textit{\emph{combo}} correlation represents a hybrid technique since it relates the prompt emission quantity $E_{\rm p}$, inferred from the GRB $\gamma$-ray spectrum, with the observables from the X-ray afterglow light curve, such as the rest-frame $0.3$--$10$~keV plateau luminosity $L_0$, its rest-frame duration $\tau$, and the late power-law decay index $\alpha$ \citep{Izzo2015}.
Its expression is given by
\begin{equation}
\label{Comborel}
\log \left(\frac{L_0}{{\rm erg/s}}\right) = q_0 + q_1 \log \left(\frac{E_{\rm p}}{{\rm keV}}\right) - \log \left(\frac{\tau/{\rm s}}{|1+\alpha|}\right)\ ,
\end{equation}
where $q_0$ and $q_1$ are the two calibration constants. The correlation is characterized by an extra scatter: $\sigma_{\rm q}$.


\section{Calibration of GRB correlations}

Calibrating the above correlations permits us to use GRBs as distance indicators \citep{2019ApJ...873...39W,2020arXiv200303387M,2020ApJ...888...99W}. Clearly, we cannot consider GRBs as  standard candles, although we aim to alleviate the circularity problem by adopting model-independent
calibrations by means of data set that is independent enough of $H(z)$,
for which the luminosity distance $d_{\rm L}$ is not computed by assuming an \emph{\emph{a priori}} cosmological model.

In this paper, we consider a robust method for calibrating GRB correlations that we proposed in \citet{2019MNRAS.486L..46A}. In particular, we propose a \emph{\emph{\textit{\emph{"differential age method"}}.}} Spectroscopic measurements of the age difference $\Delta t$ and redshift difference $\Delta z$ of couples of passively evolving galaxies imply that $\Delta z/\Delta t\equiv dz/dt$. Under the hypothesis that galaxies form at the same redshift, the Hubble rate can be written as $H(z)=-\left(1+z\right)^{-1}\Delta z/\Delta t$. So, using the updated sample of $31$ Hubble data points, we approximate $H(z)$ using model-independent B\'ezier parametric curves.

\begin{figure}
\centering
\includegraphics[width=\hsize,clip]{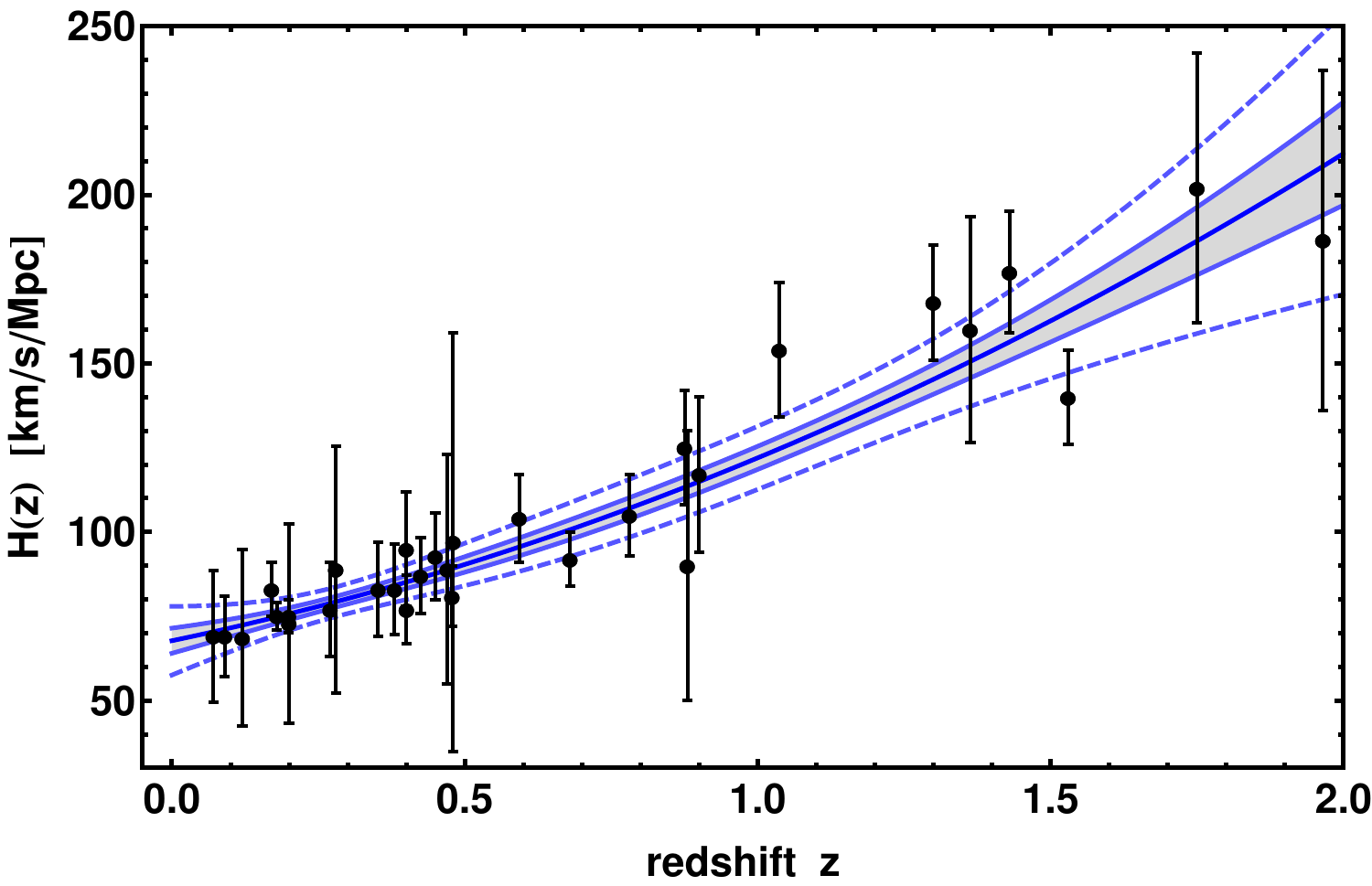}
\caption{The $H_2(z)$ function (solid thick blue) fitting OHD data (black points), with the $1\sigma$ (blue curves and shaded area) and the $3\sigma$ (blue dashed curves) confidence regions. Reproduced from \citet{2019MNRAS.486L..46A}.}
\label{fig:1}
\end{figure}

These curves are stable at the lower degrees of control points. They can be rotated and translated by performing the operations on the points and assuming a degree $n$. They are model-independent reconstructions of a  given function, and they formally read:
\begin{equation}
\label{bezier}
H_n(z)=\sum_{d=0}^{n} \beta_d h_n^d(z)\quad,\quad
h_n^d(z)\equiv \frac{n!(z/z_{\rm m})^d}{d!(n-d)!} \left(1-\frac{z}{z_{\rm m}}\right)^{n-d}\,.
\end{equation}
The coefficients $\beta_d$ of the linear combination of Bernstein basis polynomials $h_n^d(z)$ are positive in the range of $0\leq z/z_{\rm m}\leq1$, where $z_{\rm m}$ is the maximum $z$ of our Hubble data set.
Aside from cases of constant ($n=0$) and linear growth ($n=1$) and oscillatory behaviors ($n>2$), the only possible combination of B\'ezier polynomials leading to a nonlinear monotonic growing function up to $z_{\rm m}$ is $n=2$, for which the Hubble function writes as
\begin{equation}
\label{bezier2}
H_2(z)=\beta_0\left(1-\frac{z}{z_{\rm m}}\right)^2 + 2 \beta_1 \left(\frac{z}{z_{\rm m}}\right)\left(1-\frac{z}{z_{\rm m}}\right) + \beta_2 \left(\frac{z}{z_{\rm m}}\right)^2\,,
\end{equation}
where, in units of km~s$^{-1}$~Mpc$^{-1}$, we have $\beta_0=67.76\pm3.68$, $\beta_1=103.3\pm11.1$, and $\beta_2=208.4\pm14.3$.
Our results are portrayed in Fig.~\ref{fig:1}, where $H_2(z)$ is compared to the OHD data points.

\begin{figure*}
\centering
\includegraphics[width=0.49\hsize,clip]{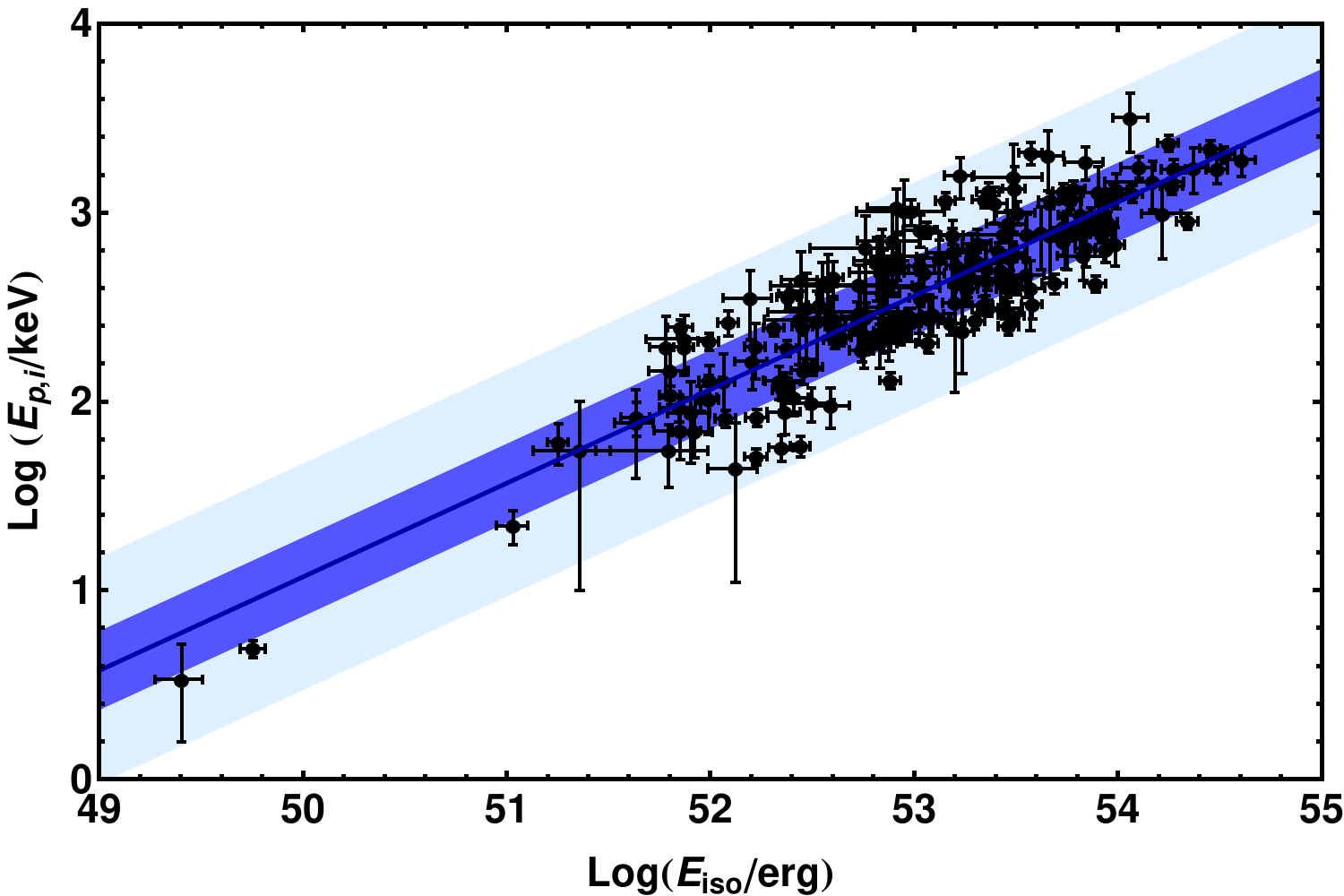}
\includegraphics[width=0.49\hsize,clip]{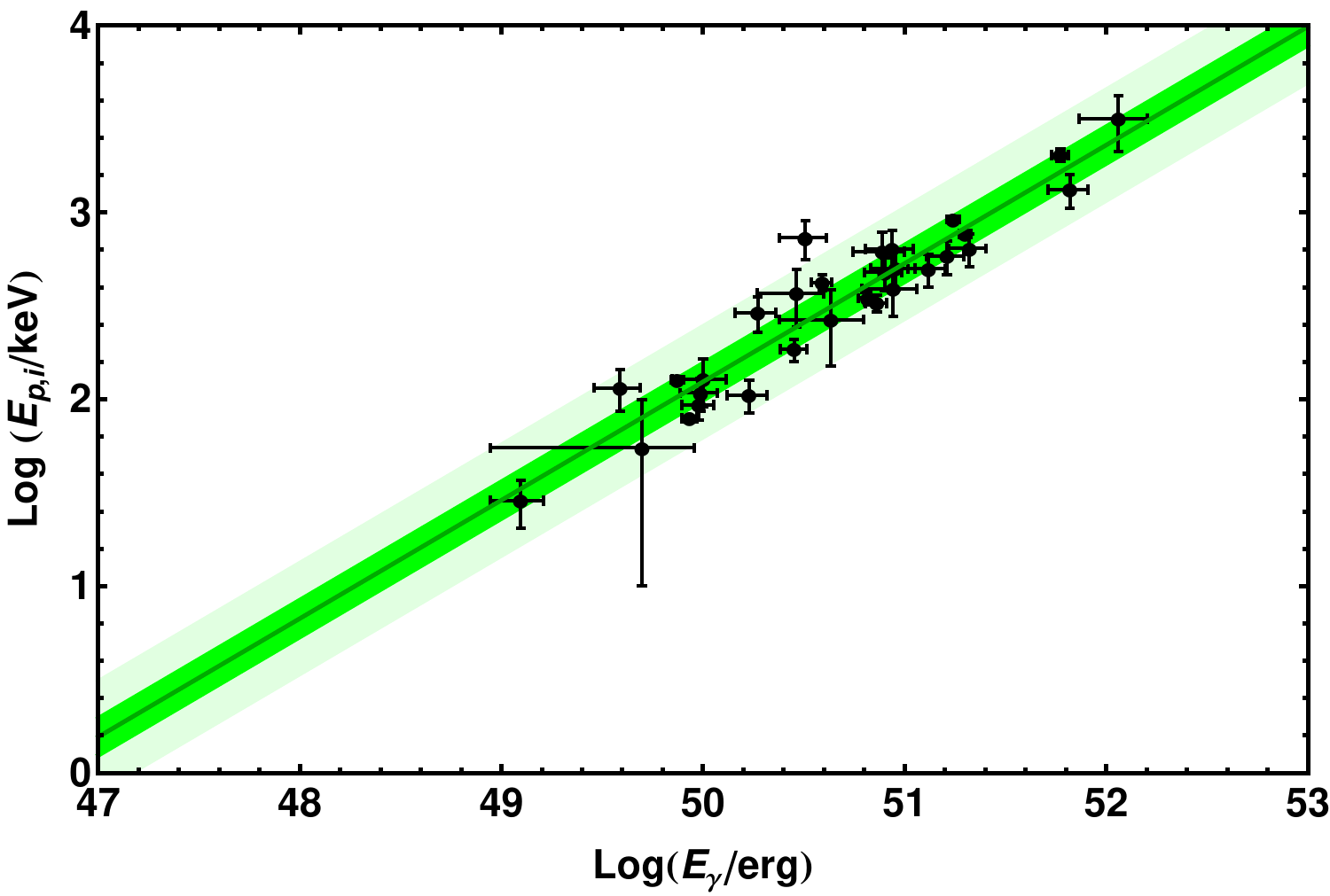}
\includegraphics[width=0.49\hsize,clip]{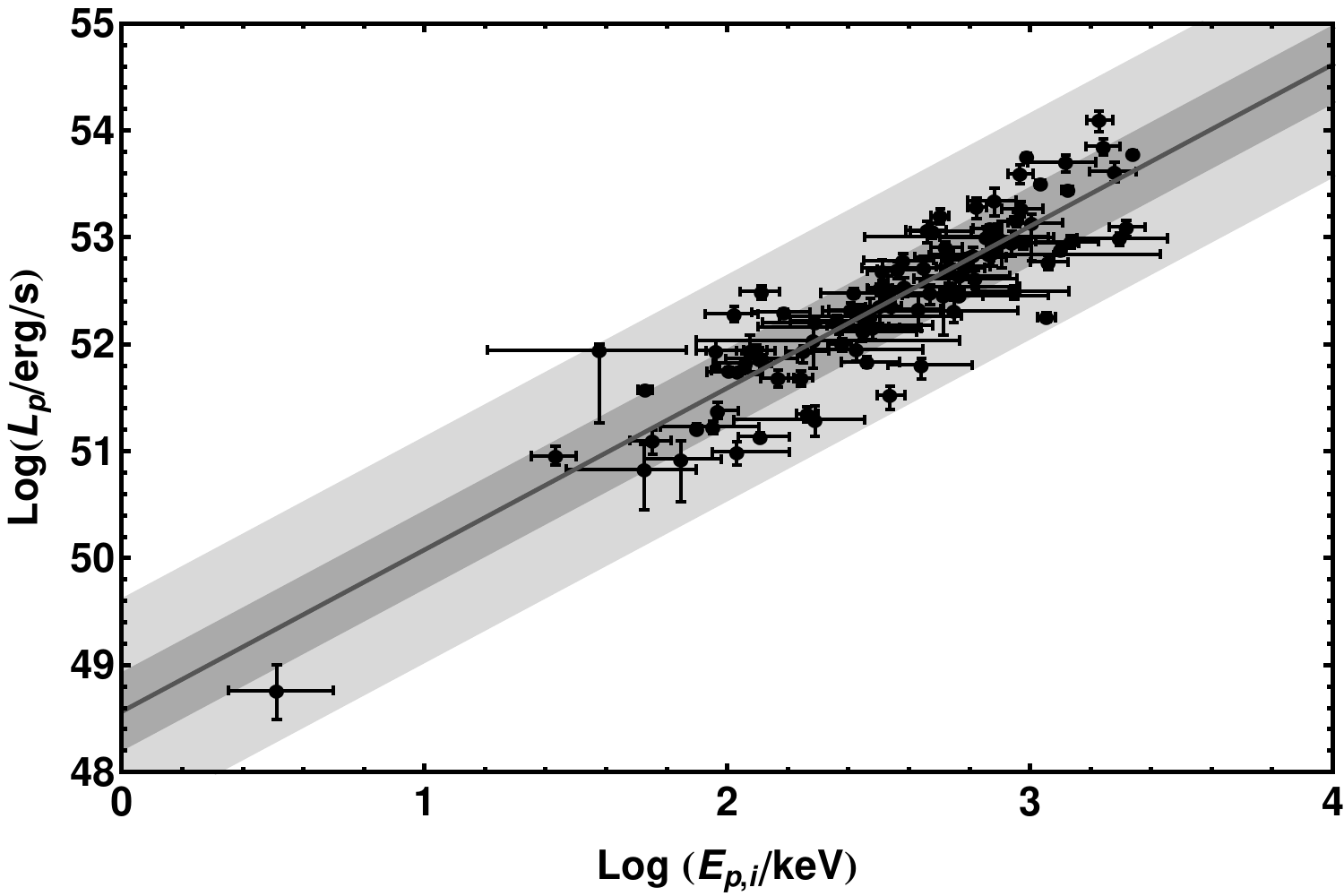}
\includegraphics[width=0.49\hsize,clip]{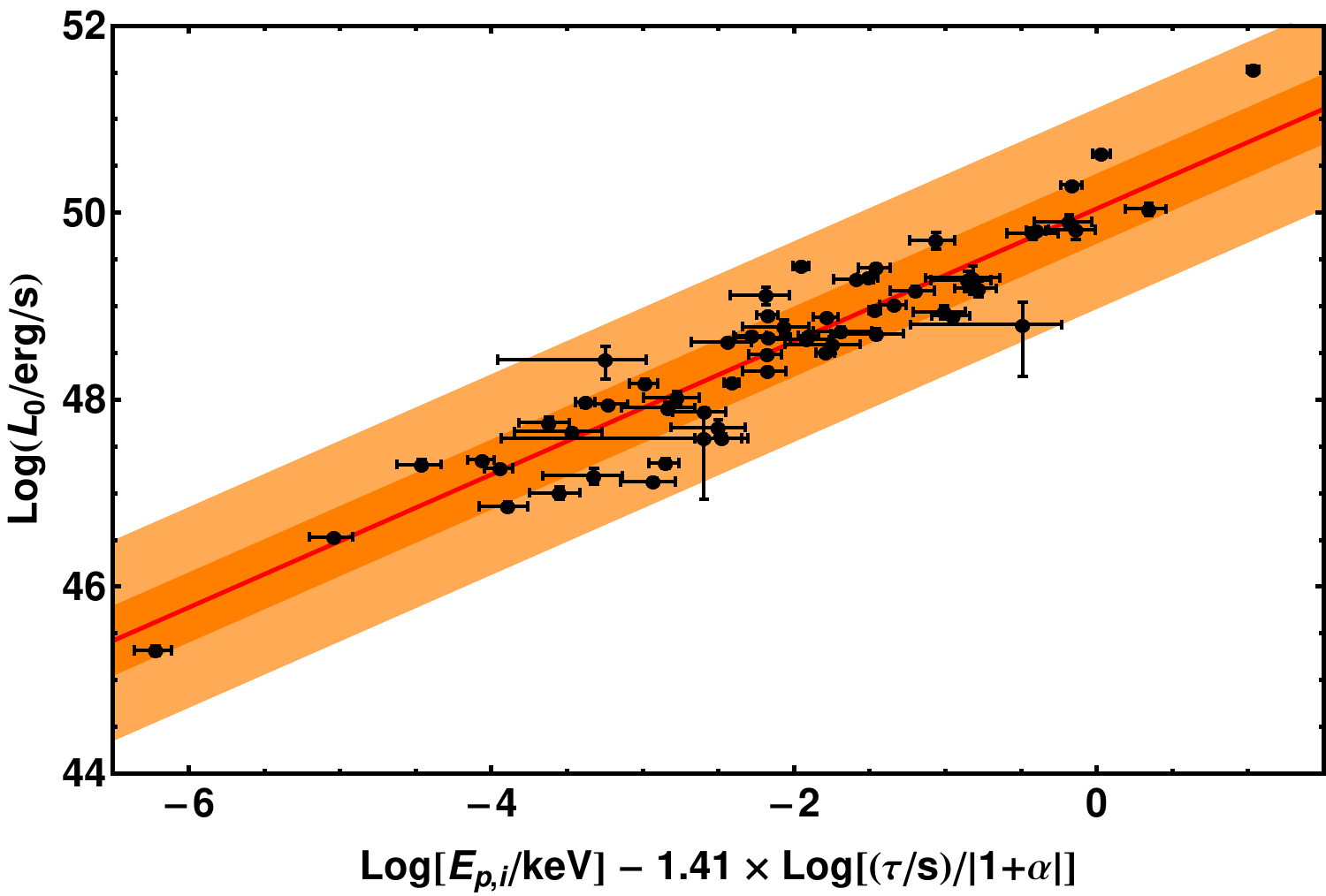}
\caption{Best fits (solid lines), their $1\sigma$ (dark shaded areas) and $3\sigma$ (light shaded areas) dispersions, and associated data sets (black points) of the calibrated GRB correlations considered in this work. Top-left: the \emph{\emph{\emph{Amati}}} correlation; top-right: the \emph{Ghirlanda} correlation; bottom-left: the \emph{Yonetoku} correlation; bottom-right: the \emph{combo} correlation.}
\label{fig:2}
\end{figure*}

After having approximated $H(z)$ with Eq.~\eqref{bezier2}, we impose a curvature parameter $\Omega_k=0$ in agreement with current indications on spatial curvature made by  \citet{Planck2018}, so that the luminosity distance becomes
\begin{equation}
\label{dlHz2}
d_{\rm cal}(z)=c(1+z)\int_0^z\dfrac{dz'}{H_2(z')}\,,
\end{equation}
enabling the calibration of $E_{\rm iso}$, $E_\gamma$, $L_{\rm p}$, and $L_0$ in the above GRB correlations and the determination of their calibration constants.
Results are shown in Fig.~\ref{fig:2} and summarized in Table~\ref{tab:1}.
\begin{table*}
\setlength{\tabcolsep}{1.5em}
\renewcommand{\arraystretch}{1.1}
\centering
\caption{Calibrated correlations. Columns: correlation, data set $N$, year of the last update, and the calibrated best fit parameters.}
\begin{tabular}{lcc|lll}
\hline\hline
Correlation         &   $N$     &   Update  & \multicolumn{3}{c}{Parameters}\\
\hline
\textit{Amati}      &   $193$   &   $2015$  &   $a_0=2.06\pm0.03$
                                            &   $a_1=0.50\pm0.02$
                                            &   $\sigma_{\rm a}=0.20\pm0.01$\\
\textit{Ghirlanda}  &   $27$    &   $2007$  &   $b_0=2.09\pm0.04$
                                            &   $b_1=0.63\pm0.04$
                                            &   $\sigma_{\rm b}=0.10\pm0.02$\\
\textit{Yonetoku}   &   $101$   &   $2018$  &   $m_0=-3.43\pm0.21$
                                            &   $m_1=1.51\pm0.08$
                                            &   $\sigma_{\rm m}=0.35\pm0.03$\\
\textit{Combo}      &   $60$    &   $2015$  &   $q_0=50.04\pm0.27$
                                            &   $q_1=0.71\pm0.11$
                                            &   $\sigma_{\rm q}=0.35\pm0.04$\\
\hline\hline
\end{tabular}
\label{tab:1}
\end{table*}

From the calibrated correlations, we are now in a position to compute their corresponding distance moduli from the standard definition: $\mu_{\rm GRB}^{\rm obs}=25+5\log(d_{\rm cal}/{\rm Mpc})$.
\begin{itemize}
\item For the \emph{Amati} correlation, from Eq.~\eqref{Amatirel}, we obtain
\begin{align}
\nonumber
\mu_{\rm A}^{\rm obs} = 32.55 + \frac{5}{2}&\left[\frac{1}{a_1}\log \left(\frac{E_{\rm p}}{{\rm keV}}\right) -\frac{a_0}{a_1} + \right.\\
\label{muGRBA}
&\left.-\log\left(\frac{4\pi S_{\rm bolo}}{\textnormal{erg/cm$^2$}}\right) +\log\left(1+z\right)\right]\,,
\end{align}
where the number in the formula takes into account the normalization of $E_{\rm iso}$ and the conversion from Mpc to cm units.
\item For the \emph{Ghirlanda} correlation, from Eq.~\eqref{Ghirlandarel}, we obtain
\begin{align}
\nonumber
\mu_{\rm G}^{\rm obs} = 27.55 + \frac{5}{2}&\left[\frac{1}{b_1}\log \left(\frac{E_{\rm p}}{{\rm keV}}\right) -\frac{b_0}{b_1} + \right.\\
\label{muGRBG}
&\left.-\log\left(\frac{4\pi\,f\,S_{\rm bolo}}{\textnormal{erg/cm$^2$}}\right) +\log\left(1+z\right)\right]\,,
\end{align}
where the number in the formula takes into account the normalization of $E_\gamma$ and the conversion from Mpc to cm units.
\item For the \textit{\emph{Yonetoku}} correlation, from Eq.~\eqref{Yonetokurel}, we obtain
\begin{align}
\nonumber
\mu_{\rm Y}^{\rm obs} = 32.55 + \frac{5}{2}&\left[m_0+m_1\log \left(\frac{E_{\rm p}}{{\rm keV}}\right) + \right.\\
\label{muGRBY}
&\left.-\log\left(\frac{4\pi F_{\rm p}}{\textnormal{erg/cm$^2$/s}}\right) +\log\left(1+z\right)\right]\,,
\end{align}
where the number in the formula takes into account the normalization of $L_{\rm p}$ and the conversion from Mpc to cm units.
\item For the \textit{\emph{combo}} correlation, from Eq.~\eqref{Comborel}, we write
\begin{align}
\nonumber
\mu_{\rm C}^{\rm obs} = -97.45 + \frac{5}{2}&\left[q_0 + q_1\log \left(\frac{E_{\rm p}}{{\rm keV}}\right) -\log\left(\frac{\tau/{\rm s}}{|1+\alpha|}\right) + \right.\\
\label{eq:no3}
&\left. -\log\left(\frac{4\pi F_0}{\textnormal{erg/cm$^2$/s}}\right)
\right]\,,
\end{align}
where the luminosity distance is expressed in cm.
\end{itemize}
The above four GRB Hubble diagrams can be employed for the fits described in the following sections.

\subsection{Connection with cosmokinematics}

By redefining $a(t)$ in terms of the redshift, $z=a^{-1}-1$ and employing the cosmological principle only, one can relate the Hubble parameter to the luminosity distance $d_{\rm L}(z)$ by
\begin{equation}
H(z)=\left\{\frac{d}{dz}\left[\frac{d_{\rm L}(z)}{1+z}\right]\right\}^{-1}\,,
\end{equation}
which results in
\begin{align}
\label{dlandH}
d_L(z)&=\frac{z}{H_0}\sum_{n=0}^{N}\frac{\alpha_n}{n!}z^n\,,\\
H(z)&=\sum_{m=0}^{M}\frac{H^{(m)}}{m!}z^m\,.
\end{align}
By construction, fixing a given expansion order $N$ in $d_L$ means that the corresponding one for $H$ is $N-1$. Thus, taking $N=4$, more explicitly, we have
\begin{equation}
d_{\rm L}^{(4)}\simeq\frac{z}{H_0}\left(\alpha_0+\alpha_1z+\alpha_2\frac{z^2}{2}+\alpha_3\frac{z^3}{6}\right)\,,
\end{equation}
or in terms of $H:$
\begin{equation}
H^{(3)}\simeq H_0\left(1+H_1z+H_2\dfrac{z^2}{2}+H_3\dfrac{z^3}{6}\right)\,.
\end{equation}
The derivatives that enter as coefficients $\alpha_i, H_i$ are the cosmographic sets that can be directly compared with cosmic data and  observations. Thus, we formally have:
\begin{align}\label{dlH}
\alpha_i&\equiv\alpha_i(q_0,j_0,s_0)\,,\\
H_i&\equiv H_i(q_0,j_0,s_0)\,,
\end{align}
where up to the fourth order in $d_L$, the cosmographic set, $q_0,j_0,s_0$, is immediately provided by current values of the derivatives of $a(t)$ by:
\begin{align}
\label{equazionissime}
H(t) &= + \frac{1}{a} \; \frac{d a}{d t}\,,\\
q(t) &= - \frac{1}{a} \; \frac{d^2 a}{d t^2}  \;\left[ \frac{1}{a} \;
\frac{d a}{d t}\right]^{-2}\,,\\
j(t) &= + \frac{1}{a} \; \frac{d^3a}{d t^3}  \; \left[ \frac{1}{a}
\; \frac{d a}{d t}\right]^{-3}\,,\\
\label{equazionissime4}
s(t) &= + \frac{1}{a} \; \frac{d^4a}{d t^4}\left[ \frac{1}{a}
\; \frac{d a}{d t}\right]^{-4}\,,
\end{align}
often referred to as cosmographic parameters. The above terms are the \emph{\emph{Hubble deceleration}}, \emph{\emph{jerk}}, and \emph{\emph{snap}} parameters, respectively \citep{2007CQGra..24.5985C,PhysRevD.78.063501}.

Cosmography is the part of cosmology that, using those coefficients, aims to reconstruct the Universe's kinematics in a model-independent way. For these reasons, we very often refer to this approach as cosmokinetics. Equations~\eqref{equazionissime}--\eqref{equazionissime4} lead to errors that are produced by  truncation, and mathematically the convergence decreases when additional higher order terms are involved. This fact clearly compromises the overall accuracy, and, motivated by this issue, we need to study different sets of parameters with fixed orders. An example would be to employ particular \emph{\emph{hierarchies}} among coefficients. We describe the hierarchical procedure in the next sections, and we highlight how cosmokinematics can be formulated at low and high redshifts by means of different reformulations of the redshift variable.


\subsection{Low-redshift kinematics}

The luminosity distance  can be connected to the cosmographic series by directly computing the coefficients that enter the previous expansions given by Eq.~\eqref{dlH}. We supposed above that the Universe is spatially flat, meaning that it has a vanishing curvature $(k=0)$ and that the cosmological principle holds. This leads to the fact that we did not consider the energy momentum budget of the Universe or its fundamental compositions. We are, however, interested in writing up the coefficients in Eq.~\eqref{dlH}, taking $\alpha_0=1,$ and
\begin{align}
\alpha_1&=\frac{1}{2}(1-q_0)\,,\\
\alpha_2&= -\frac{1}{6}(1-q_0-3q_0^2 +j_0)\,,\\
\alpha_3&=\dfrac{1}{24}(2 - 2 q_0 - 15 q_0^2 - 15 q_0^3 + 5 j_0 +10 q_0 j_0 +s_0)\,,
\end{align}
and
\begin{align}
H_1&=1 + q_0\,,\\
H_2&=\dfrac{1}{2} (j_0 - q_0^2)\,, \\
H_3&=\dfrac{1}{6}\left[-3 q_0^2 - 3 q_0^3 + j_0 (3 + 4 q_0) + s_0\right]\,,
\end{align}
corresponding to the direct derivative of the functions involved, meaning $d_{\rm L}$ and $H$, with respect to time.


\subsection{High-redshift cosmography}

Although powerful, the above formalism suffers from shortcomings due to the convergence at higher redshifts. The violation of $z\simeq0$ implies that series lose their meanings because they are computed at our time only. In other words, the standard cosmographic approach fails to be predictive if one employs data at higher redshift domains, which is exactly the case of GRBs. It is, in principle, inconvenient to obtain information on the evolution of cosmographic coefficients when one uses data exceeding the previous redshifts. What is trickier with the approaches to high-redshift cosmography is that one has to write up new  techniques to rephrase the definition of cosmological distances.

To further analyze this severe restriction, one can perform a direct high-redshift cosmography by inventing smaller convergence radii and/or theoretical series that are different from Taylor ones, making it possible to heal the \emph{\emph{problem of convergence}} arising for cosmography at high redshifts \citep{2020arXiv200309341C}.

In summary, the strategies used to address the issues of high-redshift cosmography are, so far: 1) extending the  limited convergence radii of Taylor series by changing variables of expansion \citep{2019IJMPD..2830016C} and 2) changing the way in which the expansions are performed, through smooth functions, typically ratios of polynomials, which better converge at higher redshift \citep{2014PhRvD..90d3531A}. Below, we analyze the first and second approaches in detail and develop the use of auxiliary variables and rational approximations, respectively.

In the case of auxiliary variables, one employs a tricky method in which the expansion variable is reformulated, under the condition that the new variable satisfies a peculiar construction obeying a few mathematical conditions. The idea is commonly to restrict the convergence radius up to a given value, substituting $z$, which goes up to infinity, with a new variable that does not.
In such a way, the convergence radius is severely reduced in a new Taylor series (reformulated in terms of an independent variable): hereafter $y\equiv\mathcal{F}(z)$.
All prototypes are functions of the redshift $z$ and obey the following requirements as a minimum:
\begin{itemize}
    \item[{\bf 1.}]  $\mathcal F(z)\Big|{\,}_{z=0}=0$\,,
    \item[{\bf 2.}]  $\mathcal F(z)\Big|{\,}_{z=0}<\infty$\,.
\end{itemize}
A more suitable $\mathcal F(z)$ might address the additional requirement:
\begin{itemize}
    \item[{\bf 3.}] $\mathcal F(z)\Big|{\,}_{z=-1}<\infty$\,.
\end{itemize}
Following this formula, we considered a first variable  under the form $y_1=z/(1+z)$, satisfying the first two conditions above.  Although appealing, $y_1$  was categorically ruled out for its inconsistencies in fitting cosmic data \citep[for details, see, e.g.,][]{2015PhRvD..92l3512B,2018MNRAS.476.3924C,2018JCAP...07..037R}. Thus, a second, more refined one that even includes the further condition \textbf{3.} can be written as
\begin{equation}
y_2=\arctan{z}\,.
\end{equation}
The second one has shown to better perform cosmographic analyses, and, in fact, satisfies the whole set of conditions \emph{\emph{in toto}}, since $y_2(0)=0$ and $y_2(\infty)=\pi/2$.

We can compute the coefficients of Eq.~\eqref{dlH}, using $y_2$ by
\begin{align}
\alpha_1&=\dfrac{1}{2}(1 - q_0)\,,\\
\alpha_2&= \dfrac{1}{6} (1 - j_0 + q_0 + 3 q_0^2)\,,\\
\alpha_3&=\dfrac{1}{24}(10 + 5 j_0 - 10 q_0 + 10 j_0 q_0 - 15 q_0^2- 15 q_0^3 + s_0)\,,
\end{align}
and
\begin{align}
H_1&=1 + q_0\,,\\
H_2&=\dfrac{1}{2} (j_0 - q_0^2)\,, \\
H_3&=\dfrac{1}{6} \left[2 + 2 q_0 + 3 q_0^2 + 3 q_0^3 - j_0 (3 + 4 q_0) - s_0 \right]\,.
\end{align}

Next, we discuss rational approximations. Another way to extrapolate information from improved versions of the series is to extend the convergence radius by changing the way in which expansions are effectively performed. To do so, one can consider rational approximations, meaning well-established prescriptions in extending Taylor expansions in case of a numerical pathology of slopes.

Optimizing the Taylor series with rational approximants leads to the construction of new polynomials that approach infinity at larger $z$ better than the Taylor series. This procedure guarantees mathematical stability of the new series if data points exceed the low redshift limit and better adapts to combined surveys involved in cosmographic computations. Among all the possible choices, Pad\'e and Chebyshev polynomials represent outstanding examples that have been proposed so far \citep{2018MNRAS.476.3924C}. In this work, we consider the Pad\'e polynomials, firstly introduced in \citep{2014PhRvD..89j3506G}. This technique of approximations turns out to be a bookkeeping device to keep the calculations manageable for the cosmography convergence issue.

The only disadvantage of rational approximations is that the orders of Pad\'e expansions are, in principle, not known. So, exploiting the correct expansion orders becomes of fundamental interest in developing our analyses, although the price to pay is that error bars commonly increase with respect to standard Taylor approaches. Intense investigations have been performed throughout recent years \citep[see, e.g.,][]{2016EPJC...76..281Z}. To do so, one can estimate the cosmographic parameters and the corresponding error bars circumscribing the analyses to the series that better provide  stable results. We therefore limit our analyses below to $P_{2,1}$ Pad\'e  approximations. Thus, provided we have Taylor expansions of $f(z)$ under the form $
f(z)=\sum_{i=0}^\infty c_i z^i$, with  $c_i=f^{(i)}(0)/i!$, it is possible to obtain the $(n,m)$ Pad\'e approximant of $f(z)$ by
\begin{equation}
P_{n,m}(z)=\left(\sum_{i=0}^{n}a_i z^i\right)\left(1+\sum_{j=1}^{m}b_j z^j\right)^{-1}\,,
\label{eq:def Pade}
\end{equation}
which, by construction, is rational and requires that $b_0=1$. Furthermore, it is important that $
f(z)-P_{n,m}(z)=\mathcal{O}(z^{n+m+1})$ and the coefficients $b_i$ come from  solving the homogeneous system of linear equations $
\sum_{j=1}^m b_j\ c_{n+k+j}=-b_0\ c_{n+k}$,
valid for $k=1,\hdots,m$. Once $b_i$ are known, $a_i$ can be obtained using the formula
$a_i=\sum_{k=0}^i b_{i-k}\ c_{k}$.
Finally, we get
\begin{equation}
P_{2,1}(z)=\frac{z}{H_0}\left\{\frac{6 (q_0-1) + \left[q_0 (8 + 3 q_0) -5 - 2 j_0\right] z}{-2 (3 + z + j_0 z) + 2 q_0 (3 + z + 3 q_0 z)}\right\}\,.
\end{equation}

We take into account low- and high-redshift cosmographic methods, and below we show how to obtain results from Monte Carlo procedures using the two aforementioned hierarchies.


\section{Methods of statistical analysis}

Our cosmographic fits utilize  high-redshift data from GRBs. In particular, the points are in the form of Hubble diagram data, obtained from the calibrated correlations described by the numerical coefficients prompted in Table~\ref{tab:1}.
To perform our fits, we considered Markov chain Monte Carlo (MCMC) simulations sampled from the widest possible parameter space by means of a modified version of the freely available \texttt{Wolfram Mathematica} code provided in \citet{2019PhRvD..99d3516A}.
We employed the widely adopted Metropolis-Hastings algorithm. We can therefore compare our approach with several others developed in the literature, and we reduce the dependence on initial statistical distributions made in the simulations by means of the properties of the algorithm itself. The numerical procedure, in particular, is summarized in detail in what follows below for the data sets composed by supernovae, BAO, and GRBs.


\subsection{Numerical procedures}

We computed the best fit cosmological parameters, namely ${\bf x}$, by minimizing the total $\chi^2$ function, built up from the GRB, SNe Ia, and BAO data sets. We define it as
\begin{equation}
\chi^2_{\rm tot}=\chi^2_{\rm GRB}+\chi^2_{\rm SN}+\chi^2_{\rm BAO}\,.
\end{equation}
Below, we define the $\chi^2$ for each of the considered probes.

\begin{itemize}
\item For each of the considered GRB correlations, we define
\begin{equation}
\label{chisquared}
\chi^2_{\rm GRB}=\sum_{i=1}^{N_{\rm GRB}}\left[\dfrac{\mu_{\rm GRB,i}^{\rm obs}-\mu_{\rm GRB}^{\rm th}\left({\bf x},z_i\right)}{\sigma_{\mu_{\rm GRB,i}}}\right]^2\,,
\end{equation}
where $N_{\rm GRB}$ is summarized in Table~\ref{tab:1}, $\mu_{\rm GRB}^{\rm th}$ are the theoretical GRB distance moduli for a given model.
\item For SNe Ia, we computed the $\chi^2$ function from the \textit{\emph{Pantheon sample}}, which is the largest and most recent combined sample consisting of $1048$ SNe Ia ranging in $0.01<z<2.3$ \citep{2018ApJ...859..101S}.
The SNe Ia distance modulus is parameterized as
\begin{equation}
\mu_{\rm SN}=m_{\rm B}- \left(\mathcal{M}-\alpha \mathcal{X}_1+\beta \mathcal{C} -\Delta_{\rm M}-\Delta_{\rm B}\right)\ ,
\end{equation}
where $m_{\rm B}$ and $\mathcal{M}$ are the $B$-band apparent and absolute magnitudes, respectively. Furthermore, $\mathcal{X}_1$ and $\mathcal{C}$ are the SN light-curve stretch and the color factors, respectively; $\alpha$ and $\beta$ are the coefficients of the luminosity-stretch and luminosity-color relationships respectively.
Finally, $\Delta_{\rm M}$ is a distance correction based on the host galaxy mass of SNe, and $\Delta_{\rm B}$ is a distance correction based on predicted biases from simulations.

The uncertainties of each SN do not depend on $\mathcal{M}$, therefore, the $\chi^2$ of the SN data is given by
\begin{equation}
\chi^2_{\rm SN}=\left(\Delta \mathbf{\mu}_{\rm SN}- \mathcal{M}\mathbf{1} \right)^{\rm T} \mathbf{C}^{-1}
\left(\Delta\mathbf{\mu}_{\rm SN}-\mathcal{M} \mathbf{1} \right)\,,
\end{equation}
where $\Delta\mu_{\rm SN}\equiv \mu_{\rm SN}-\mu_{\rm SN}^{\rm th}\left({\bf x},z_i\right)$ is the module of the vector of residuals, and $\mathbf{C}$ is the covariance matrix that contains statistical and systematic uncertainties on the light-curve parameters \citep{2011ApJS..192....1C}.
The parameter $\mathcal{M}$ can be removed from the fits by analytically
marginalizing it and assuming a flat prior  \cite{2001A&A...380....6G}, leading to %
\begin{equation}
 \chi^2_{{\rm SN},\mathcal{M}} = a + \log \frac{e}{2 \pi} - \frac{b^2}{e}\,,
 \label{eqn:chimarg}
\end{equation}
where $a\equiv\Delta\vec{\mathbf{\mu} }_{\rm SN}^{T}\mathbf{C}^{-1}\Delta\vec{\mathbf{\mu} }_{\rm SN}$, $b\equiv\Delta\vec{\mathbf{\mu} }_{\rm SN}^{T}\mathbf{C}^{-1}\vec{\mathbf{1}} $, and $e \equiv
\vec{\mathbf{1}}^T\mathbf{C}^{-1} \vec{\mathbf{1}} $.
Analytical marginalizations over $\alpha$ and $\beta$ were not performed, because they contribute to the uncertainties of each SN.
\item The BAO is produced by the propagation of sound waves in the early Universe. These waves are observed as a peak in the large-scale structure correlation function and provide a measure of  angular distance. We track these waves by means of the comoving volume variation $D_{\rm V}^3({\bf x},z)$ at a given $z$, through the BAO observable for uncorrelated data $d_{\rm z}({\bf x},z)$, namely
\begin{equation}
\label{eq:DV}
D_{\rm V}^3({\bf x},z) \equiv \frac{c\,z}{H({\bf x},z)}\left[\frac{d_{\rm L}({\bf x},z)}{1+z}\right]^2\ \ \,,\ \ d_{\rm z}({\bf x},z) \equiv \frac{r_{\rm s} (z_\text{d})}{D_{\rm V}({\bf x},z)}\,.
\end{equation}
However, BAO points are slightly model dependent.
The comoving sound horizon $r_{\rm s}(z_\text{d})$ in Eq.~\eqref{eq:DV} depends upon the baryon drag redshift $z_\text{d}$ and needs to be calibrated with CMB data for a given  cosmological model.
For the MCMC simulations, we used \citet{Planckfirstrelease} best fit values $z_\text{d}=1059.62\pm0.31$ and $r_{\rm s}(z_\text{d})=147.41\pm0.30$.

We used the BAO data shown in Table~\ref{tab:BAO}.
We did not include correlated BAO from the WiggleZ data \citep{2011MNRAS.418.1707B}, because the corresponding observable depends on the cosmological parameter $\Omega_{\rm m}$. This permits largely model-independent BAO points, through which we obtain
\begin{equation}
\chi^2_{\rm BAO}=\sum_{i=1}^{N_{\rm BAO}} \left[\frac{d_{\rm z,i}^{\rm obs}-d_{\rm z}^{\rm th}({\bf x},z_i)}{\sigma_{d_{\rm z,i}}}\right]^2\ .
\end{equation}
\end{itemize}
\begin{table}
\setlength{\tabcolsep}{0.3em}
\renewcommand{\arraystretch}{1.0}
\begin{tabular}{l c c l}
\hline\hline
Survey & $z$ & $d_{\rm z}$ & Reference\\
\hline
6dFGS & $0.106$ & $0.3360\pm0.0150$ & \citet{2011MNRAS.416.3017B}\\
SDSS-DR7 & $0.15$ & $0.2239\pm0.0084$ & \citet{2015MNRAS.449..835R}\\
SDSS &  $0.20$ & $0.1905\pm0.0061$ & \citet{2010MNRAS.401.2148P}\\
SDSS-III & $0.32$ & $0.1181\pm0.0023$ & \citet{2014MNRAS.441...24A}\\
SDSS & $0.35$ & $0.1097\pm0.0036$ & \citet{2010MNRAS.401.2148P}\\
SDSS-III & $0.57$ & $0.0726\pm0.0007$ & \citet{2014MNRAS.441...24A}\\
SDSS-III & $2.34$ & $0.0320\pm0.0016$ & \citet{2015AA...574A..59D}\\
SDSS-III & $2.36$ & $0.0329\pm0.0012$ & \citet{2014JCAP...05..027F}\\
\hline
\hline
\end{tabular}
\caption{Uncorrelated BAO data set with surveys and references.}
\label{tab:BAO}
\end{table}


\subsection{Technical procedures of numerical analyses}

In view of our numerical analyses, we considered \textit{\emph{Amati}}, \textit{\emph{Ghirlanda}}, \textit{\emph{Yonetoku}}, and \textit{\emph{combo}} correlations. For every correlation, we investigated the cosmographic approach by adopting the strategies of expanding in Taylor series first and then with $y_2$ and rational approximations. As discussed, we considered up to the first order the Taylor series, whereas we employed a $y_2$ variable and $P_{2,1}$ Pad\'e rational polynomial. In doing so, we explored the ranges of parameters reported in Table~\ref{priori}. The intervals were built up to guarantee that the minima occurring in our computations are unique and correspond to real minima in the statistical analyses. The spread over the polynomial priors corresponds to the increased complexity in analyzing a series. The difficulty of getting results from the aforementioned definitions of $z$, $y_2,$ and $P_{2,1}$ is compatible with previous efforts made in constraining the cosmological distances. We expand up to the fourth order $d_{\rm L}$ and up to the third order $d_{rm z}$. In the latter case, however, the expansion is decreased by the term $\sim z^{-1}$ that enters Eq.~\eqref{eq:DV}. Moreover, to constrain up to the jerk parameter within \textit{\emph{hierarchy 1}}, we decreased the order of expansion of a unity. The same procedure is valid even for the auxiliary variable $y_2$. For the sake of clarity, in the case of rational approximation, we notice a favorite fitting stability when including a Taylor $d_{\rm z}$ with the Pad\'e expansion of $d_{\rm L}$. Thus, a mixed version of our fitting procedure seems to provide suitable results, which means that stable numerics are obtained at the end of the computations. The strategies are compatible with what has been proposed so far in the literature in the case of rational approximations. The procedures enable us to explore the widest ranges of degrees of freedom by means of the aforementioned priors. No poles in the rational approximations occur for the spanned ranges of coefficients employed here. The numerical bounds inferred are therefore reliable for checking the goodness of any cosmological paradigm. In what follows, we give details on our numerical expectations and on their consequences in view of our understanding toward the cosmological scenario.
\begin{table}
\centering
\setlength{\tabcolsep}{0.6em}
\renewcommand{\arraystretch}{1.0}
\begin{tabular}{c c c c}
\hline\hline
Parameters & Redshift $z$ & Function $y_2$ & Pad\'e $P_{2,1}$ \\
\hline
$h_0$          & [0.4,1.0]      & [0.4,1.0] & [0.4,1.0]  \\
$q_0$          & [-1.0,0.0]     & [-1.5,0.5] & [-2.0,1.0] \\
$j_0$          & [-2.0,2.0]     & [-2.0,3.0] & [-2.0,3.0] \\
$s_0$ & [-2.0,2.0] & [-6.0,6.0] & [-6.0,6.0] \\
\hline
\hline
\end{tabular}
\caption{Priors of our numerical fits. The large intervals for $y_2$ and $P_{2,1}$ are due to the computational complexity that increases as alternative versions of expansions are involved into computation. In our results, no poles and/or divergences occur for $P_{2,1}$ in the range of explored redshifts.}
\label{priori}
\end{table}


\section{Discussing numerical estimates}
\label{results}

The strategy adopted here is to employ SNeIa, BAO ,and GRB data sets in fitting the hierarchies $\mathcal A_1$ and $\mathcal A_2$.
We did not consider using GRBs alone, although our calibrations would permit us to consider them as \emph{\emph{standardized}}. In fact, in support of our choice, if we considered GRBs alone, the final output  over $H_0$ would be unconstrained in analogy to what happens for SNe. Moreover, for the sake of clarity, the controversy over the $H_0$ measurements is therefore unsolved in this scheme, even when adding more than one data set. This occurs since the $H_0$ value, thus far obtained from the OHD-based calibration procedure, turns out to be compatible with the current estimates by the \citet{Planck2018} and in agreement at the $1.49\sigma$ level with the value measured by \citet{2018ApJ...861..126R}, respectively. Moreover, our final results indicate a non definitive concordance with $H_0$ with respect to the two accepted results over Hubble's parameter today.
\begin{table*}
\centering
\setlength{\tabcolsep}{0.22em}
\renewcommand{\arraystretch}{1.5}
\begin{tabular}{l|ccc|cccc}
\hline\hline
{\bf Taylor fits}         &  \multicolumn{3}{c}{\emph{Hierarchy} $1$}
            &  \multicolumn{4}{c}{\emph{Hierarchy} $2$}\\
\cline{2-8}
Sample      &  $h_0$
            &  $q_0$
            &  $j_0$
            &  $h_0$
            &  $q_0$
            &  $j_0$
            &  $s_0$\\
\hline
{\it Amati} &  $0.740_{-0.006\,(-0.013)}^{+0.005\,(+0.010)}$
            &  $-0.68_{-0.02\,(-0.04)}^{+0.03\,(+0.06)}$
            &  $0.77_{-0.10\,(-0.20)}^{+0.08\,(+0.16)}$
            &  $0.700_{-0.008\,(-0.015)}^{+0.007\,(+0.014)}$
            &  $-0.51_{-0.01\,(-0.02)}^{+0.02\,(+0.03)}$
            &  $0.71_{-0.05\,(-0.10)}^{+0.06\,(+0.12)}$
            &  $-0.36_{-0.10\,(-0.20)}^{+0.05\,(+0.13)}$\\
{\it Ghirlanda} &  $0.716_{-0.006\,(-0.014)}^{+0.006\,(+0.013)}$
            &  $-0.63_{-0.03\,(-0.05)}^{+0.03\,(+0.06)}$
            &  $0.76_{-0.09\,(-0.18)}^{+0.09\,(+0.17)}$
            &  $0.691_{-0.007\,(-0.015)}^{+0.008\,(+0.016)}$
            &  $-0.50_{-0.02\,(-0.05)}^{+0.02\,(+0.05)}$
            &  $0.64_{-0.10\,(-0.19)}^{+0.06\,(+0.15)}$
            &  $-0.42_{-0.08\,(-0.16)}^{+0.10\,(+0.17)}$\\
{\it Yonetoku} &  $0.737_{-0.008\,(-0.015)}^{+0.008\,(+0.014)}$
            &  $-0.73_{-0.01\,(-0.04)}^{+0.03\,(+0.06)}$
            &  $0.88_{-0.13\,(-0.23)}^{+0.02\,(+0.13)}$
            &  $0.695_{-0.008\,(-0.015)}^{+0.007\,(+0.014)}$
            &  $-0.54_{-0.01\,(-0.03)}^{+0.02\,(+0.04)}$
            &  $0.70_{-0.05\,(-0.11)}^{+0.07\,(+0.13)}$
            &  $-0.36_{-0.09\,(-0.18)}^{+0.08\,(+0.16)}$\\
{\it Combo} &  $0.706_{-0.007\,(-0.013)}^{+0.007\,(+0.013)}$
            &  $-0.59_{-0.03\,(-0.06)}^{+0.03\,(+0.07)}$
            &  $0.72_{-0.10\,(-0.18)}^{+0.09\,(+0.18)}$
            &  $0.693_{-0.009\,(-0.015)}^{+0.006\,(+0.014)}$
            &  $-0.52_{-0.01\,(-0.03)}^{+0.02\,(+0.05)}$
            &  $0.73_{-0.09\,(-0.15)}^{+0.06\,(+0.13)}$
            &  $-0.38_{-0.10\,(-0.19)}^{+0.06\,(+0.14)}$\\
\hline
\hline
\end{tabular}
\caption{Cosmographic best fits and $1$--$\sigma$ ($2$--$\sigma$) errors from Taylor expansions labeled as \textit{\emph{hierarchy} 1} ($h_0$, $q_0$, $j_0$) and \textit{\emph{hierarchy} 2} ($h_0$, $q_0$, $j_0$, $s_0$).}
\label{tab:summarytaylor}
\end{table*}
\begin{table*}
\centering
\setlength{\tabcolsep}{0.22em}
\renewcommand{\arraystretch}{1.5}
\begin{tabular}{l|ccc|cccc}
\hline\hline
{\bf $y_2$ fits}         &  \multicolumn{3}{c}{\emph{Hierarchy} $1$}
            &  \multicolumn{4}{c}{\emph{Hierarchy} $2$}\\
\cline{2-8}
Sample      &  $h_0$
            &  $q_0$
            &  $j_0$
            &  $h_0$
            &  $q_0$
            &  $j_0$
            &  $s_0$\\
\hline
{\it Amati} &  $0.76_{-0.01\,(-0.02)}^{+0.01\,(+0.02)}$
            &  $-1.35_{-0.04\,(-0.08)}^{+0.05\,(+0.09)}$
            &  $3.85_{-0.28\,(-0.50)}^{+0.25\,(+0.49}$
            &  $0.78_{-0.01\,(-0.02)}^{+0.01\,(+0.02)}$
            &  $-0.53_{-0.04\,(-0.10)}^{+0.06\,(+0.11)}$
            &  $-2.52_{-0.42\,(-0.78)}^{+0.31\,(+0.71)}$
            &  $-4.41_{-0.58\,(-1.29)}^{+1.00\,(+1.82)}$\\
{\it Ghirlanda} &  $0.75_{-0.01\,(-0.02)}^{+0.01\,(+0.02)}$
            &  $-1.04_{-0.05\,(-0.10)}^{+0.05\,(+0.10)}$
            &  $2.40_{-0.23\,(-0.47)}^{+0.24\,(+0.47)}$
            &  $0.74_{-0.01\,(-0.02)}^{+0.01\,(+0.02)}$
            &  $-0.45_{-0.06\,(-0.13)}^{+0.10\,(+0.17)}$
            &  $-2.17_{-0.61\,(-1.14)}^{+0.42\,(+0.96)}$
            &  $-3.08_{-0.56\,(-1.28)}^{+1.41\,(+2.68)}$\\
{\it Yonetoku} &  $0.75_{-0.01\,(-0.02)}^{+0.01\,(+0.02)}$
            &  $-1.05_{-0.04\,(-0.09)}^{+0.05\,(+0.10)}$
            &  $2.47_{-0.23\,(-0.49)}^{+0.22\,(0.50)}$
            &  $0.74_{-0.01\,(-0.02)}^{+0.01\,(+0.02)}$
            &  $-0.43_{-0.10\,(-0.21)}^{+0.03\,(+0.08)}$
            &  $-2.19_{-0.32\,(-0.63)}^{+0.62\,(+1.62)}$
            &  $-2.70_{-1.04\,(-1.51)}^{+0.37\,(+0.86)}$\\
{\it Combo} &  $0.75_{-0.01\,(-0.02)}^{+0.01\,(+0.02)}$
            &  $-1.01_{-0.05\,(-0.09)}^{+0.04\,(+0.9)}$
            &  $2.29_{-0.20\,(-0.40)}^{+0.23\,(+0.44)}$
            &  $0.74_{-0.01\,(-0.02)}^{+0.01\,(+0.02)}$
            &  $-0.43_{-0.09\,(-0.17)}^{+0.06\,(+0.16)}$
            &  $-2.19_{-0.38\,(-1.03)}^{+0.64\,(+1.19)}$
            &  $-2.79_{-0.82\,(-1.50)}^{+0.90\,(+2.59)}$\\
\hline
\hline
\end{tabular}
\caption{Cosmographic best fits and $1$--$\sigma$ ($2$--$\sigma$) errors from expansions with $y_2$ labeled as \textit{\emph{hierarchy} 1} ($h_0$, $q_0$, $j_0$) and \textit{\emph{hierarchy} 2} ($h_0$, $q_0$, $j_0$, $s_0$).}
\label{tab:summary2}
\end{table*}
\begin{table}
\centering
\setlength{\tabcolsep}{0.22em}
\renewcommand{\arraystretch}{1.5}
\begin{tabular}{l|ccc}
\hline\hline
{\bf Pad\'e fits}         &  \multicolumn{3}{c}{\emph{Hierarchy} $1$}\\
\cline{2-4}
Sample      &  $h_0$
            &  $q_0$
            &  $j_0$\\
\hline
{\it Amati} &  $0.70_{-0.02\,(-0.03)}^{+0.01\,(+0.03)}$
            &  $-0.33_{-0.03\,(-0.08)}^{+0.05\,(+0.09)}$
            &  $0.240_{-0.010\,(-0.020)}^{+0.010\,(+0.020)}$\\
{\it Ghirlanda} &  $0.70_{-0.01\,(-0.02)}^{+0.02\,(+0.03)}$
            &  $-0.31_{-0.05\,(-0.09)}^{+0.02\,(+0.06)}$
            &  $0.235_{-0.002\,(-0.006)}^{+0.013\,(+0.027)}$\\
{\it Yonetoku} &
$0.68_{-0.01\,(-0.02)}^{+0.01\,(+0.02)}$
            &  $-0.32_{-0.04\,(-0.07)}^{+0.02\,(+0.06)}$
            &  $0.240_{-0.005\,(-0.010)}^{+0.010\,(+0.021)}$\\
{\it Combo} &  $0.68_{-0.01\,(-0.02)}^{+0.01\,(+0.02)}$
            &  $-0.33_{-0.03\,(-0.06)}^{+0.03\,(+0.06)}$
            &  $0.244_{-0.006\,(-0.012)}^{+0.009\,(+0.019)}$\\
\hline
\hline
\end{tabular}
\caption{Cosmographic best fits and $1$--$\sigma$ ($2$--$\sigma$) errors from Pad\'e expansions labeled as \textit{\emph{hierarchy} 1}.}
\label{tab:summarypade}
\end{table}

\begin{figure*}
\centering
\includegraphics[width=0.49\hsize,clip]{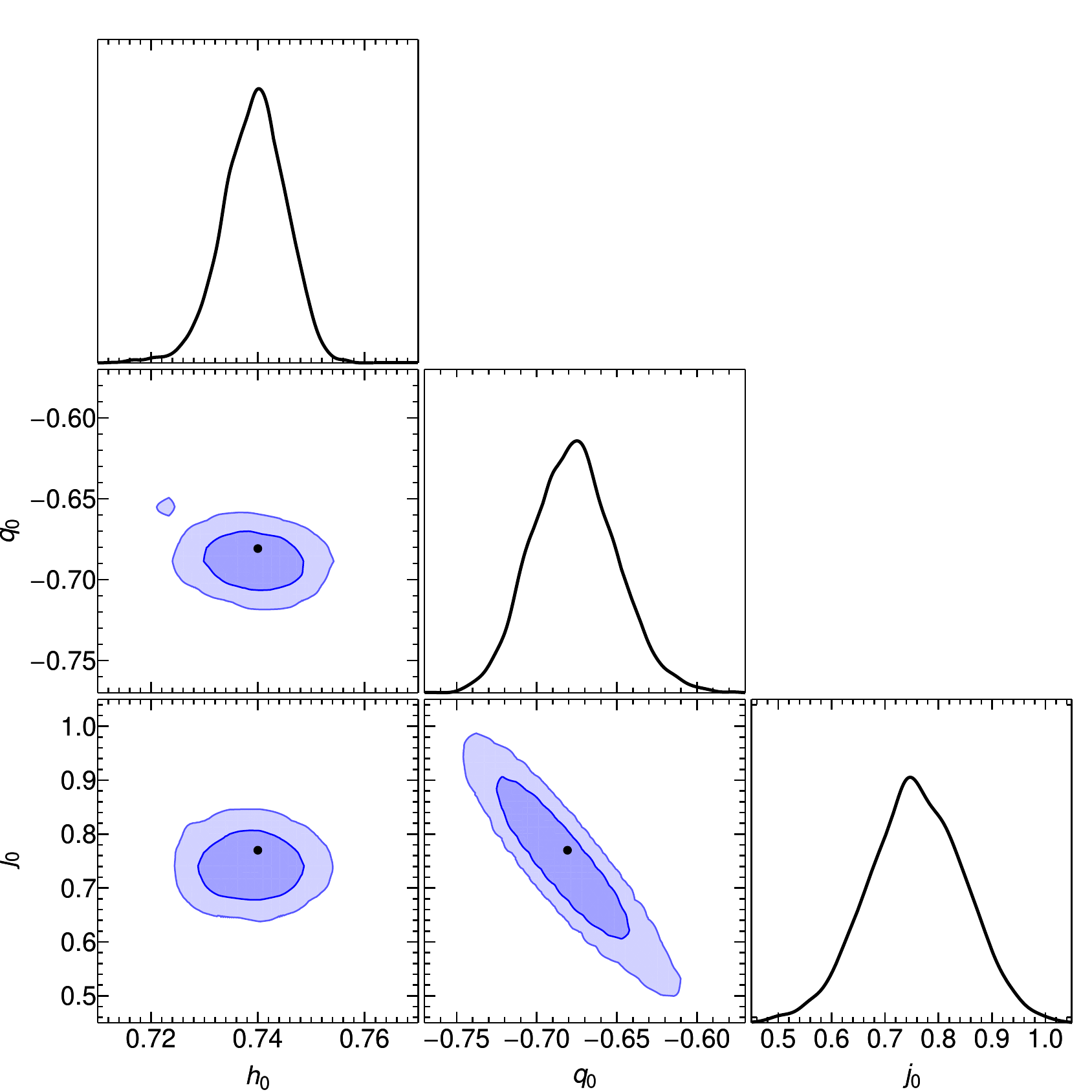}
\includegraphics[width=0.49\hsize,clip]{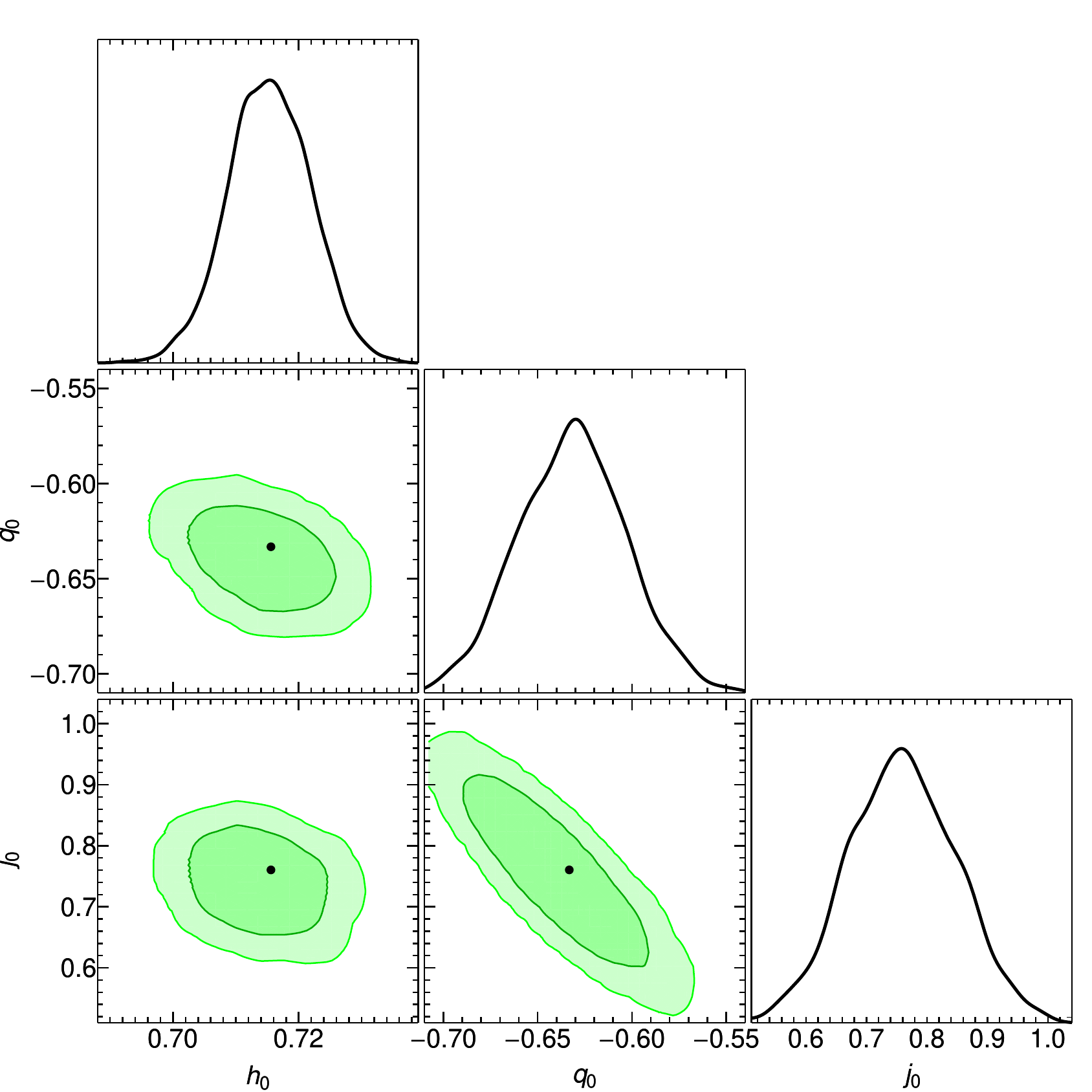}
\includegraphics[width=0.49\hsize,clip]{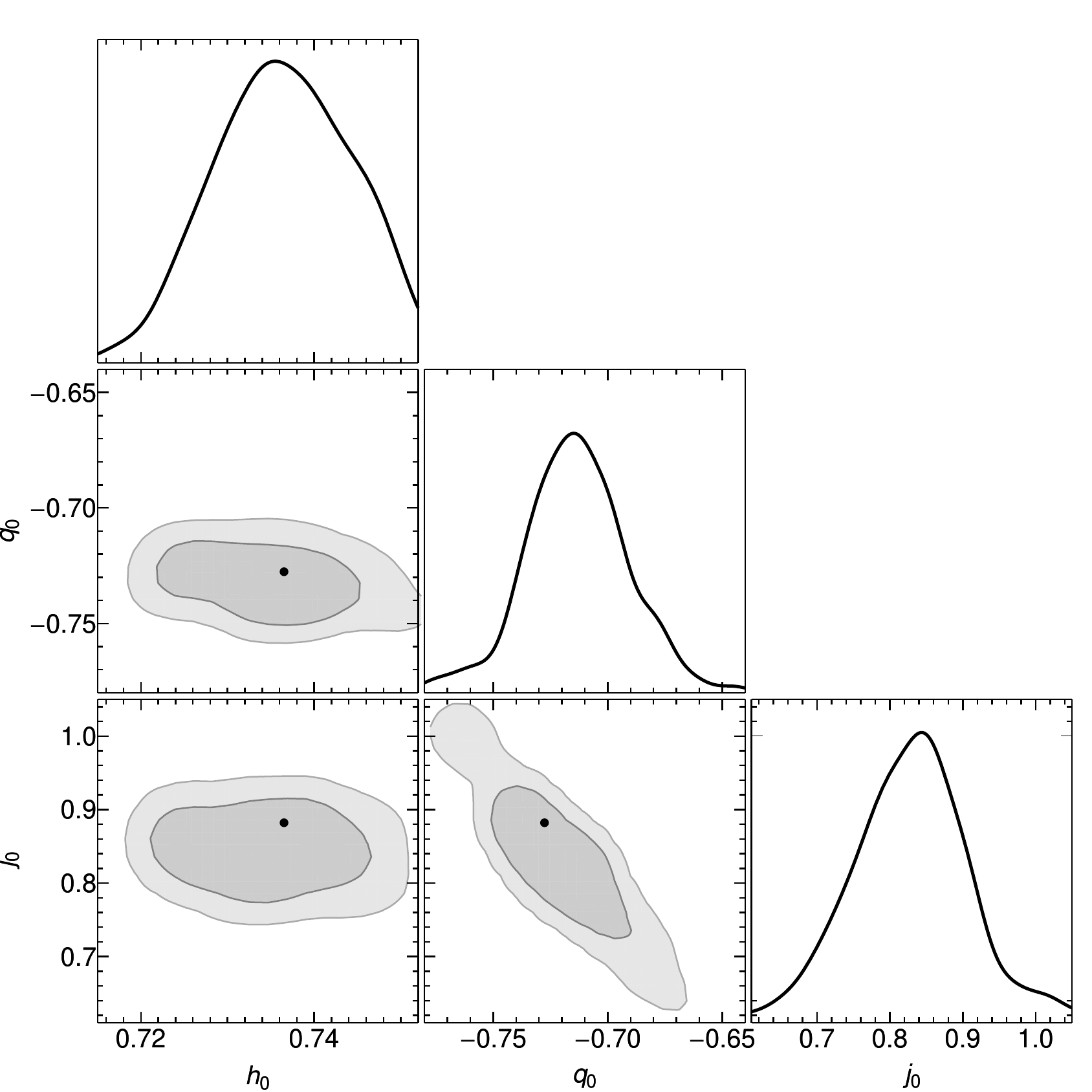}
\includegraphics[width=0.49\hsize,clip]{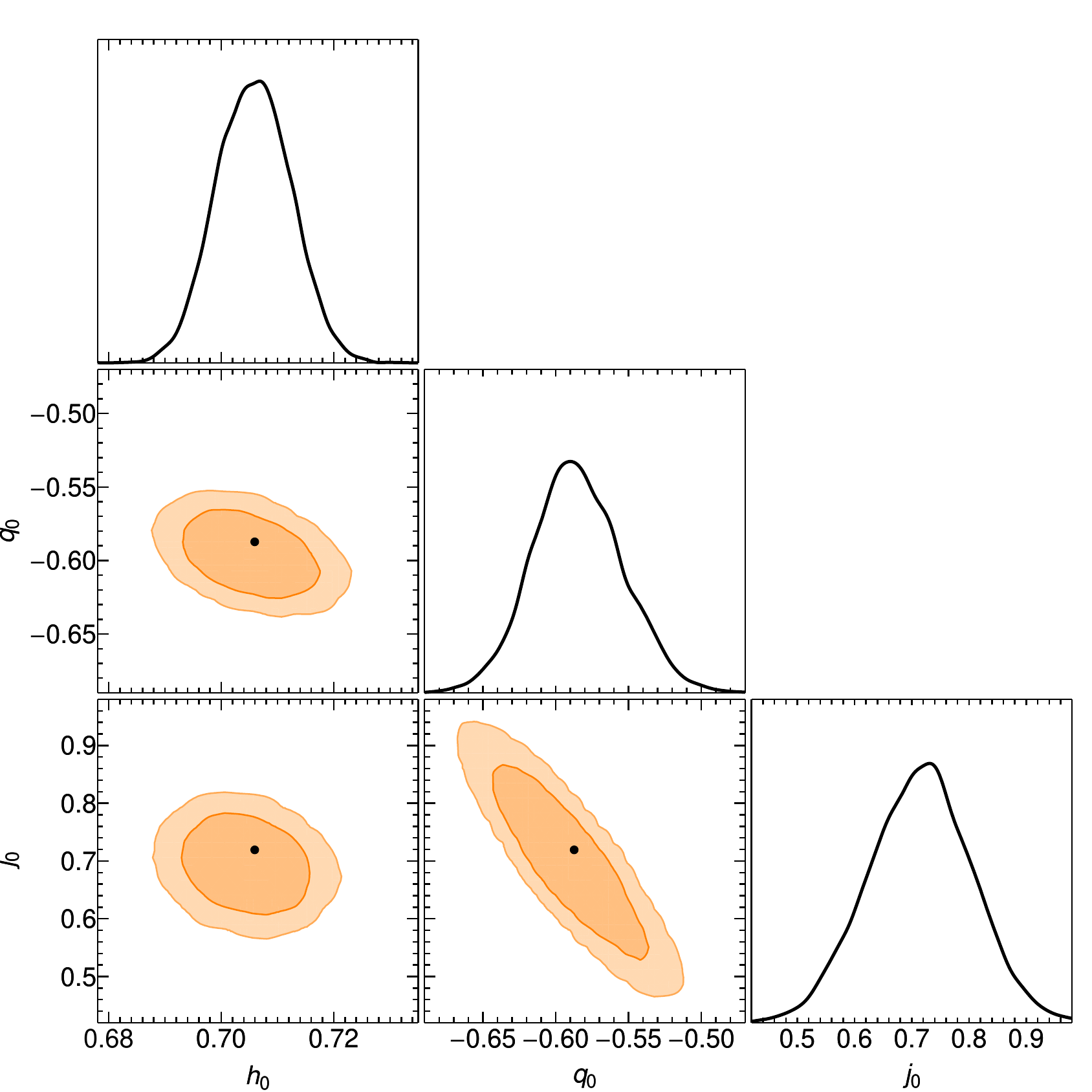}
\caption{Contours of our MCMC analyses for \textit{\emph{hierarchy} 1} Taylor expansions. Best fits (black points), prompted with $1$--$\sigma$ (darker areas) and $2$--$\sigma$ (lighter areas) confidence levels, refer to SN+BAO+GRB, in particular: \textit{\emph{Amati}} (\textit{top left}), \textit{\emph{Ghirlanda}} (\textit{top right}), \textit{\emph{Yonetoku}} (\textit{bottom left}), and \textit{\emph{combo}} (\textit{bottom right}) data sets. The results indicate a slight spread when comparing $q_0$ and $j_0$ values, with approximatively Gaussian distributions. }
\label{fig:3}
\end{figure*}

\begin{figure*}
\centering
\includegraphics[width=0.49\hsize,clip]{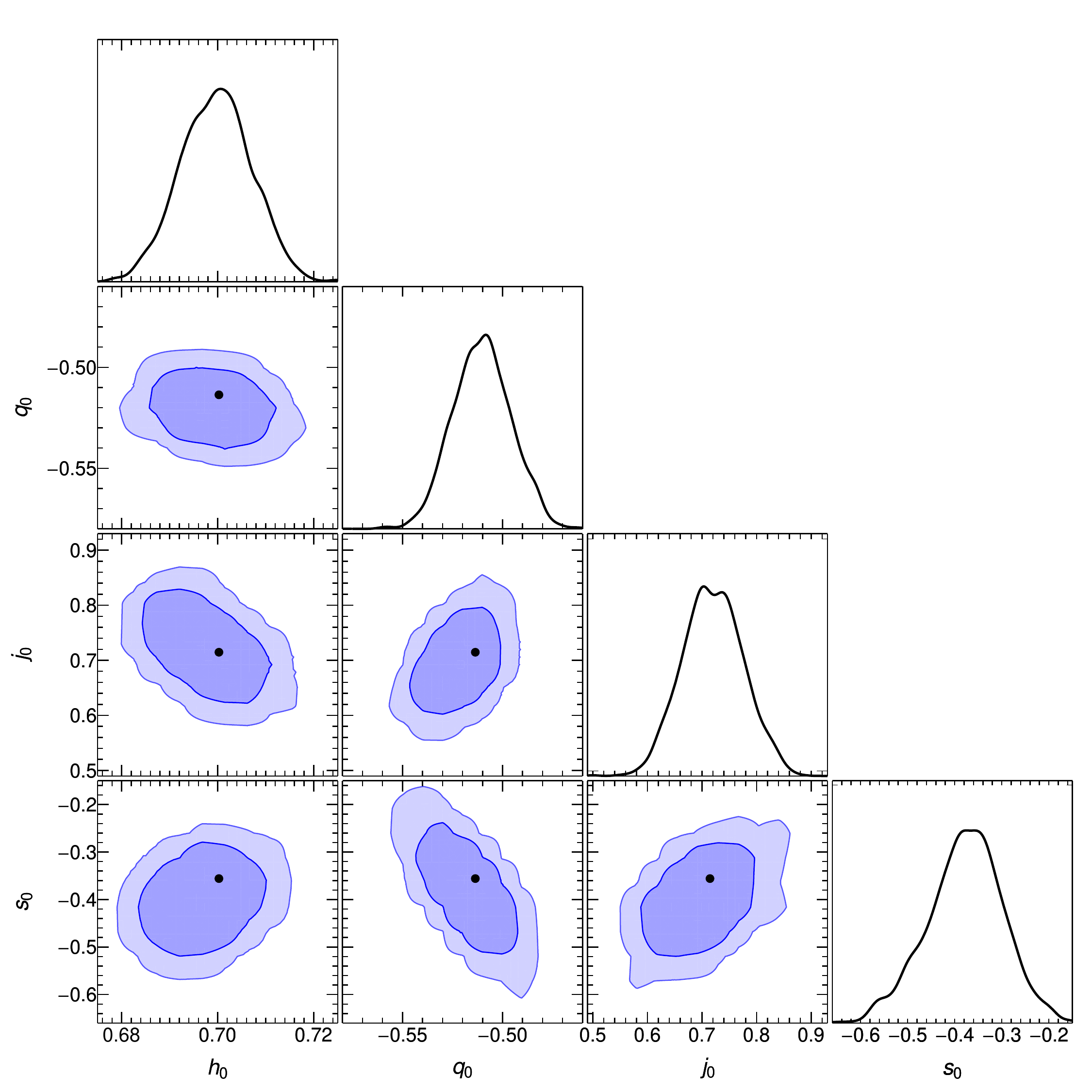}
\includegraphics[width=0.49\hsize,clip]{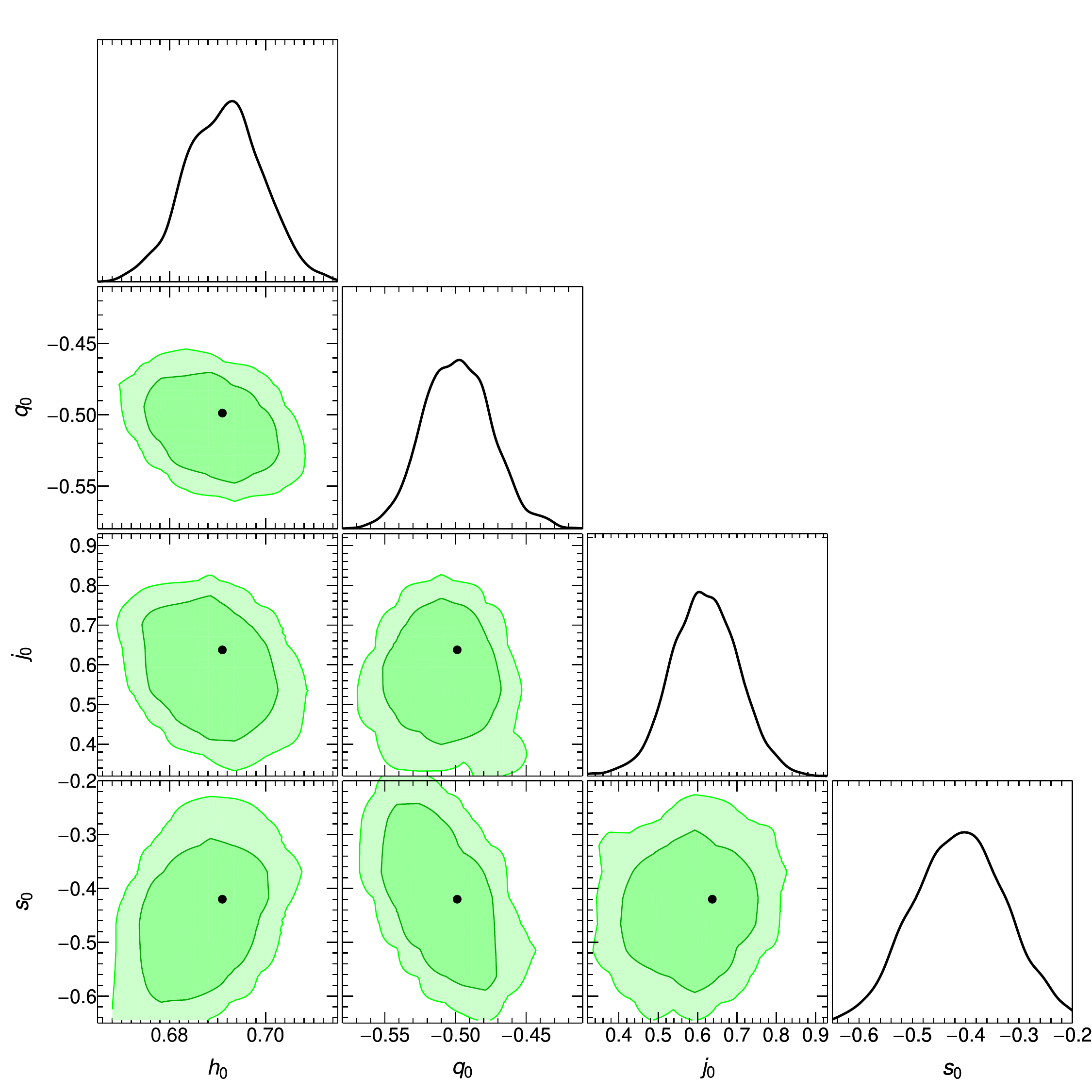}
\includegraphics[width=0.49\hsize,clip]{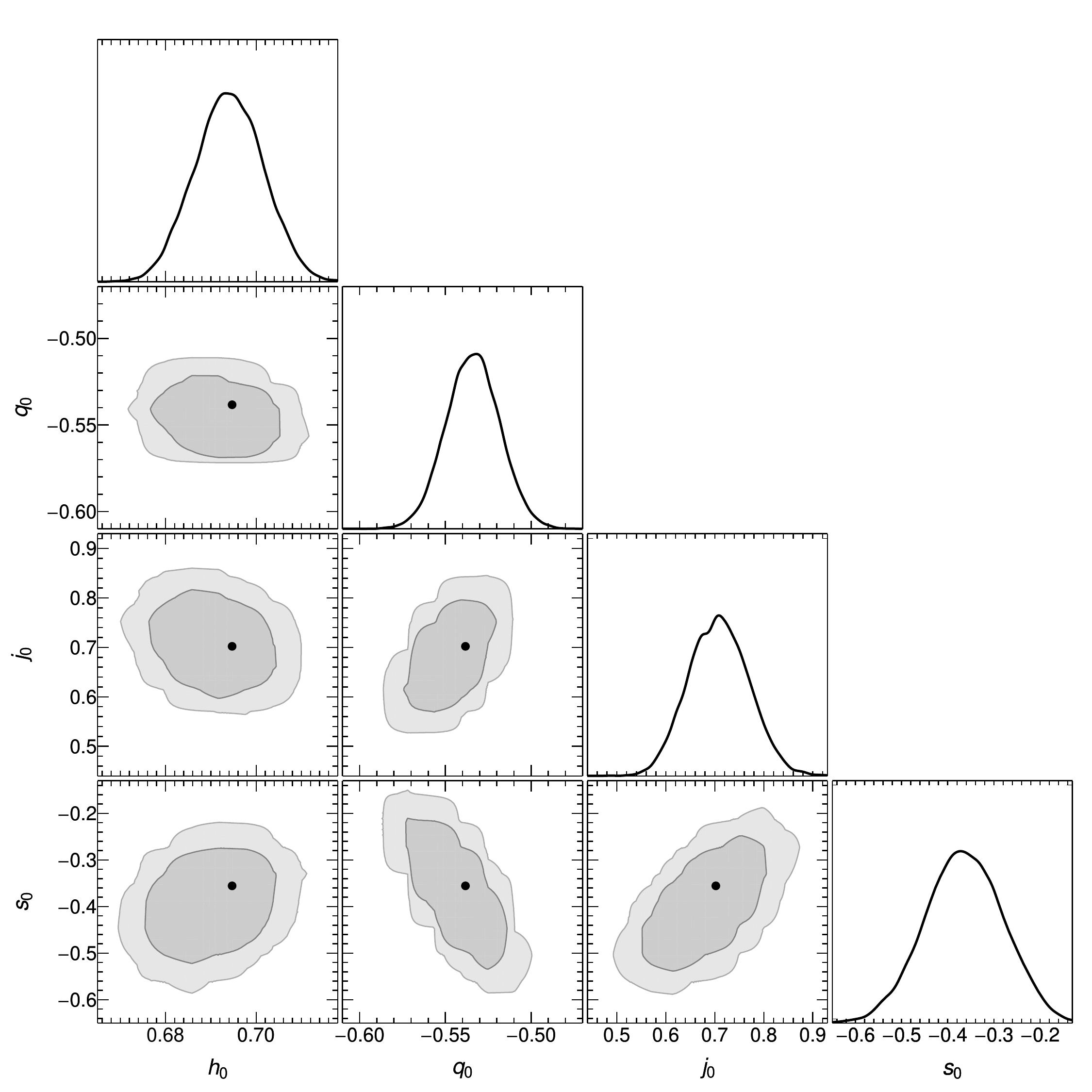}
\includegraphics[width=0.49\hsize,clip]{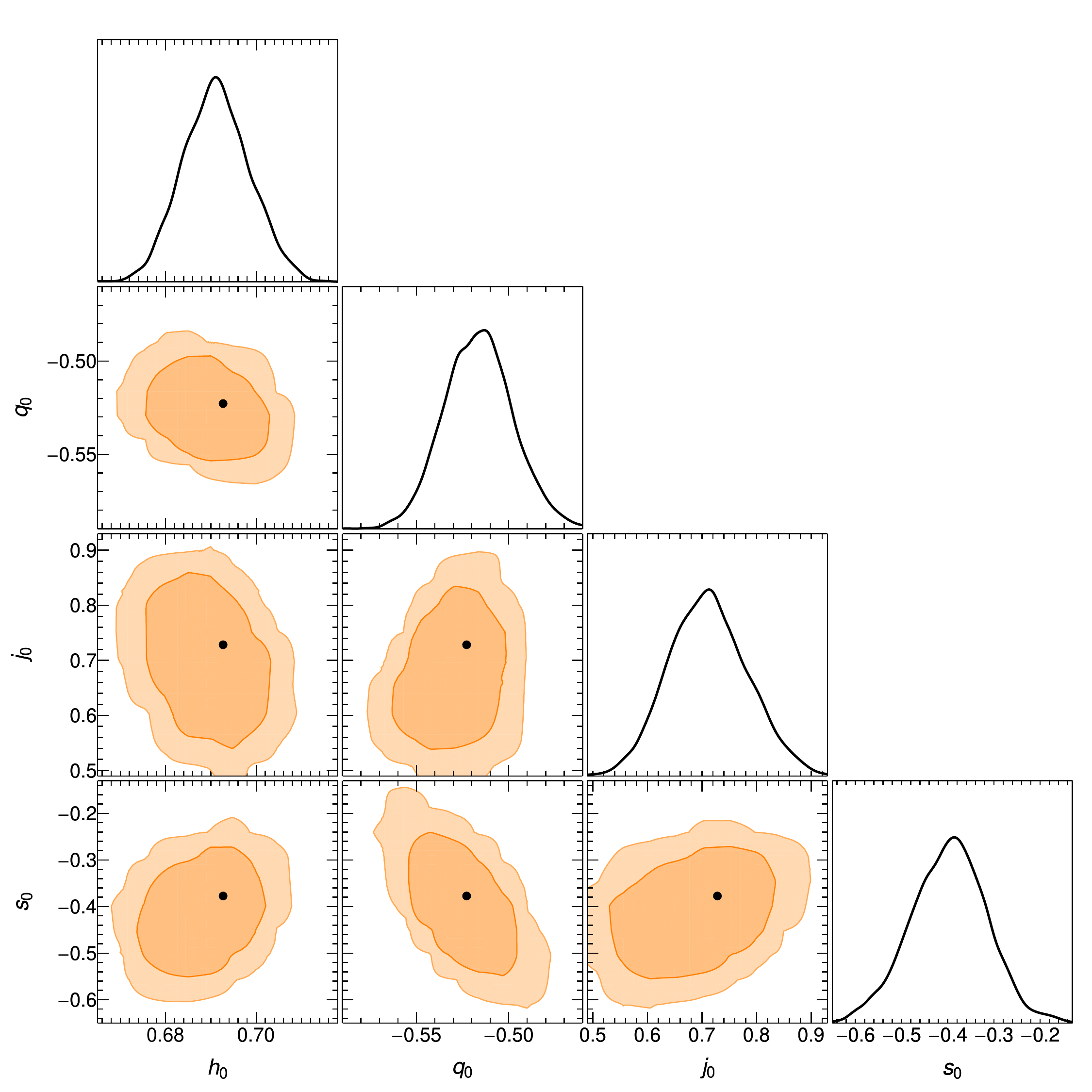}
\caption{Contours of our MCMC analyses for \emph{hierarchy} \emph{2}   Taylor expansions. Symbols, colors, and positions of the panels are the same as in Fig.~\ref{fig:3}.}
\label{fig:4}
\end{figure*}

Besides the structural form of these correlations, Taylor outcomes are effectively stable within each hierarchy, as portrayed by the results in Table~\ref{tab:summarytaylor} and Figs.~\ref{fig:3}--\ref{fig:4}, respectively. In hierarchy 1, $H_0$ values from \textit{\emph{Amati}}, \textit{\emph{Ghirlanda,}} and \textit{\emph{Yonetoku}} correlations are consistent within $1$--$\sigma$ with the \citet{2018ApJ...861..126R} prediction, meaning $(73.4\pm1.7)$~km~s$^{-1}$Mpc$^{-1}$; the  $H_0$ value from the \textit{\emph{combo}} correlation seems to be similar to \citet{Planck2018} expectations, meaning $H_0=(67.4\pm0.5)$~km~s$^{-1}$Mpc$^{-1}$, though it is still consistent with \citet{2018ApJ...861..126R} within $2$-$\sigma$. The deceleration parameter $q_0$ is compatible with a flat $\Lambda$CDM paradigm, with a mass density parameter of $\Omega_m\sim0.3$ for \textit{\emph{the} \emph{combo}} correlation, whereas the other correlations seem to indicate smaller values. In particular, \textit{\emph{the Yonetoku}} correlation indicates a value quite far from the other outcomes. Quite surprisingly, the values of $j_0$ for every correlation turns out to be incompatible with flat $\Lambda$CDM predictions. Stronger deviations are found for \textit{\emph{the combo}} correlation, meaning the same correlation for which we found $q_0$ to be in agreement with the flat concordance model in the opposing case. In hierarchy 2, Taylor fits show great concordance with \citet{Planck2018} expectations on $h_0,$ and clearly viable values over $q_0$ if compared with a flat $\Lambda$CDM model. Lower values for $j_0$ than those in hierarchy 1 are found instead. Even at $2\sigma,$ it is not possible to reach the limit $j_0=1$ prompted in the $\Lambda$CDM model. Within $1$-$\sigma,$ the snaps $s_0$ are perfectly in agreement with the flat $\Lambda$CDM paradigm, which means $s_0=-0.35$.

\begin{figure*}
\centering
\includegraphics[width=0.49\hsize,clip]{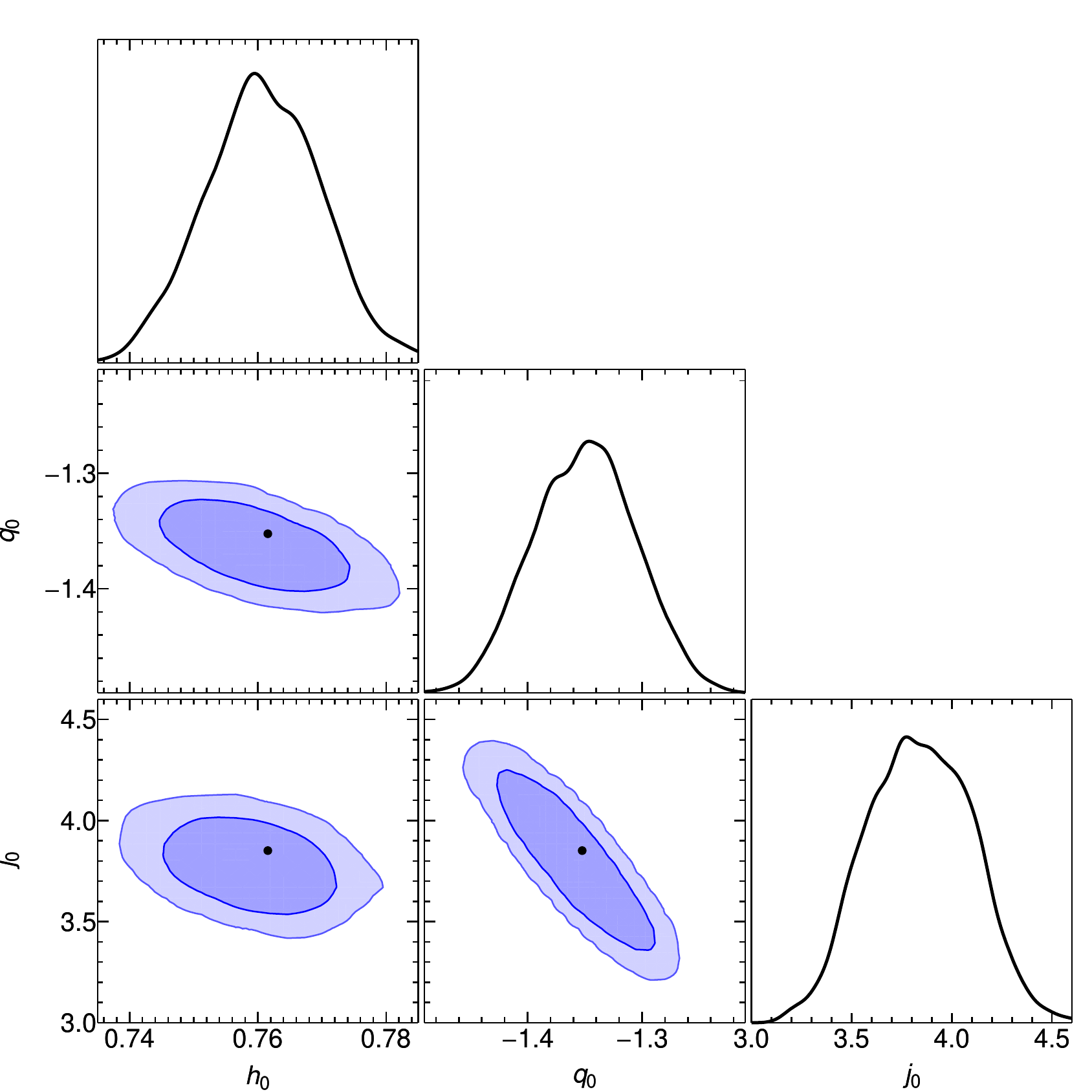}
\includegraphics[width=0.49\hsize,clip]{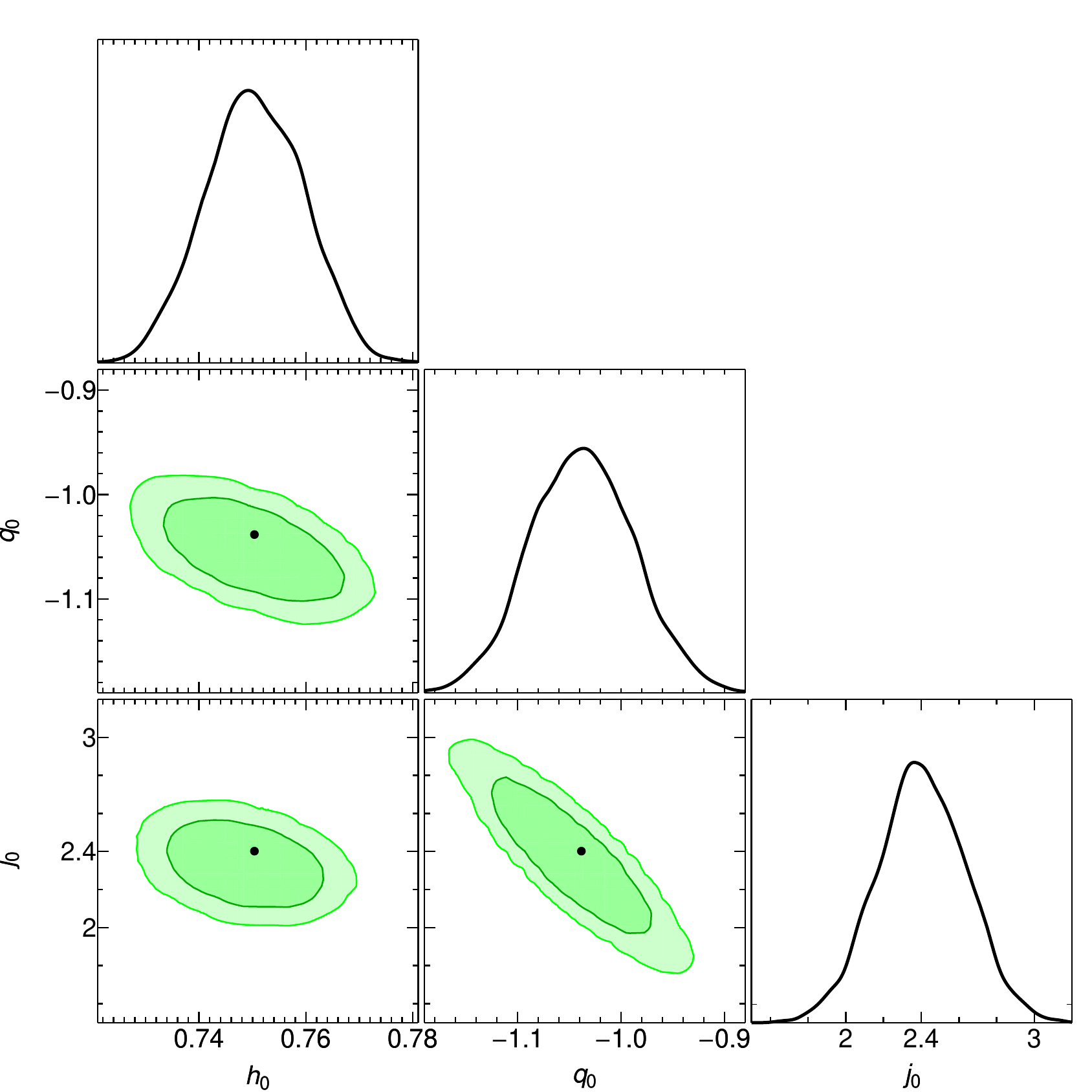}
\includegraphics[width=0.49\hsize,clip]{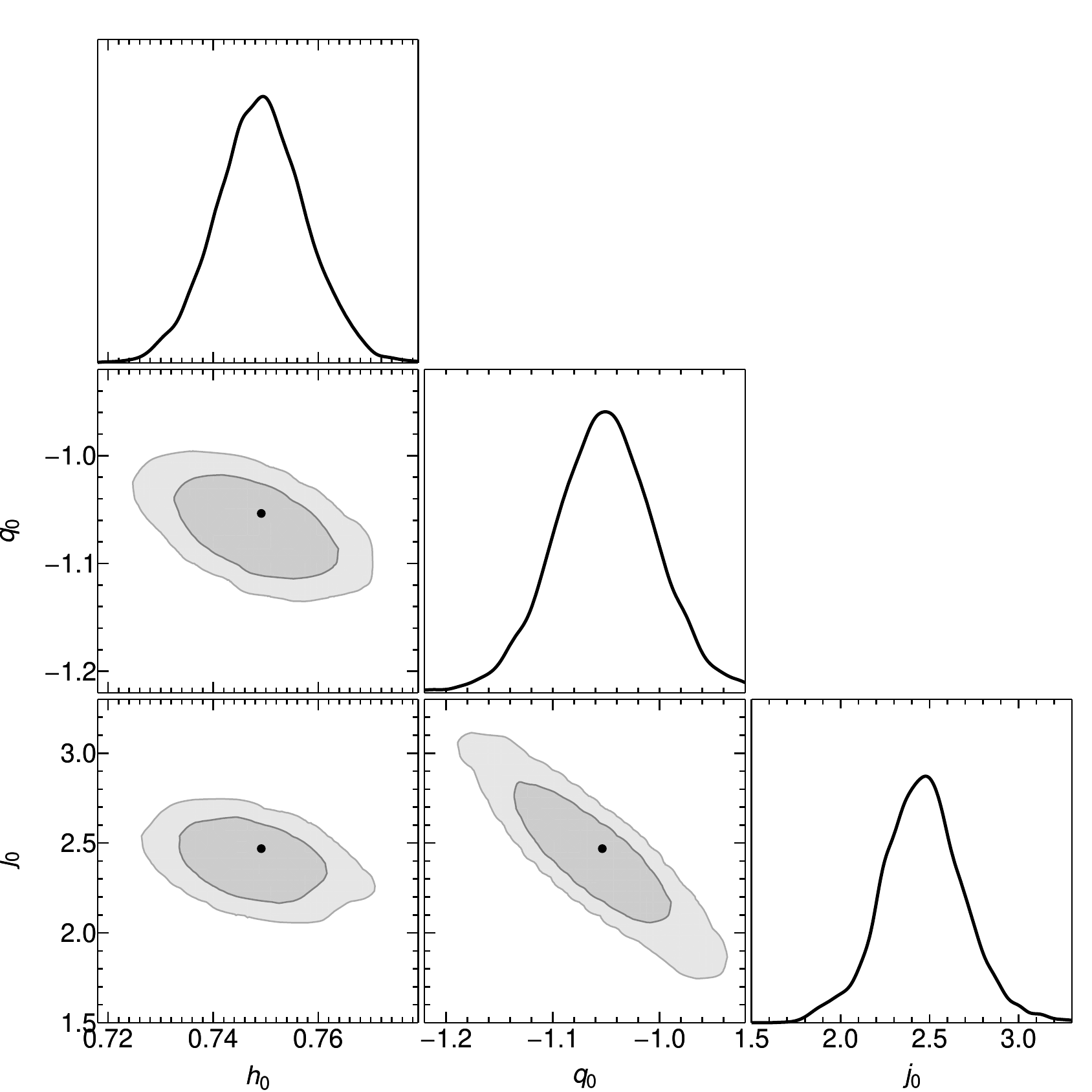}
\includegraphics[width=0.49\hsize,clip]{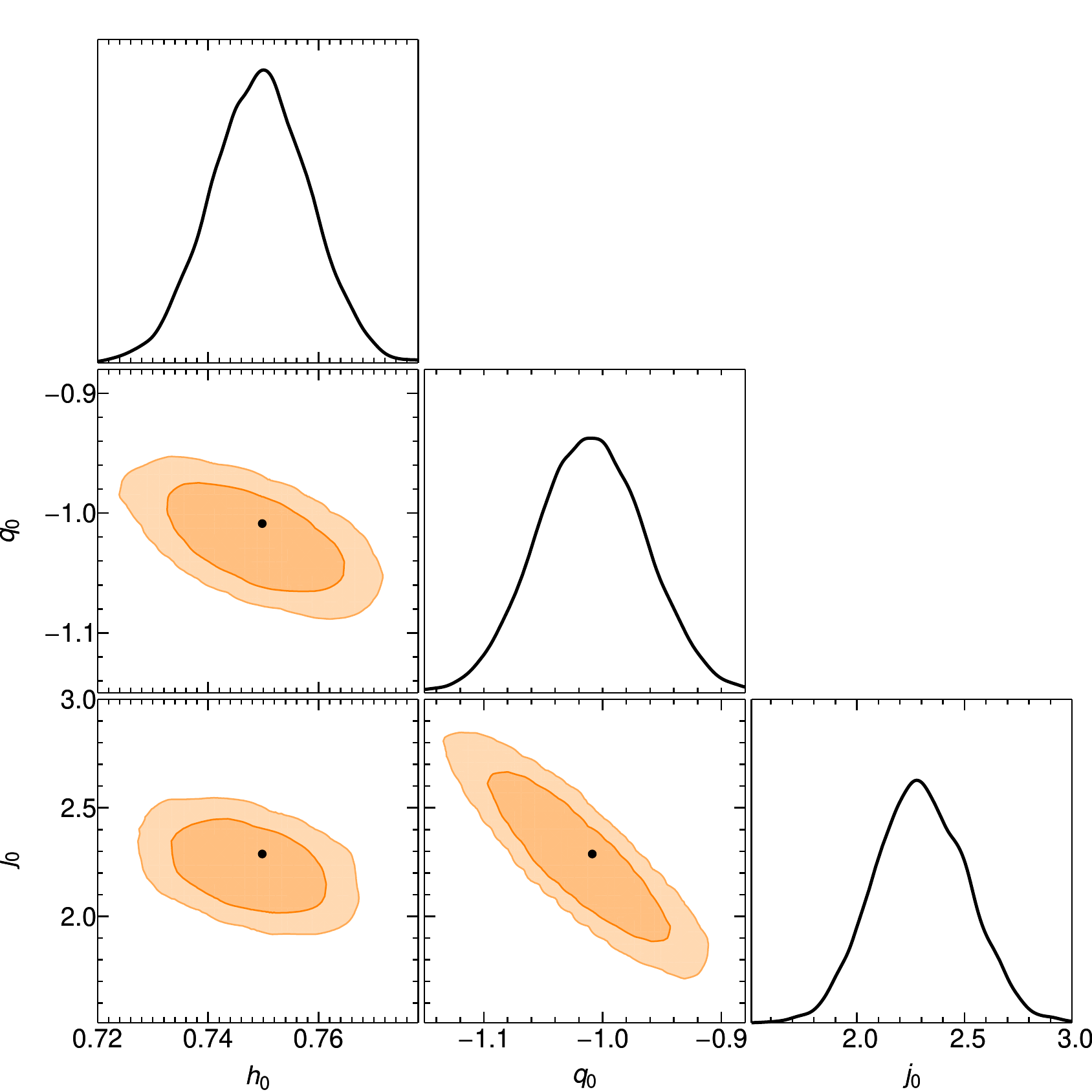}
\caption{Contours of our MCMC analyses for hierarchy 1 expansions with the auxiliary variable $y_2$. Symbols, colors, and positions of the panels are the same as in Fig.~\ref{fig:3}.}
\label{fig:6}
\end{figure*}

\begin{figure*}
\centering
\includegraphics[width=0.49\hsize,clip]{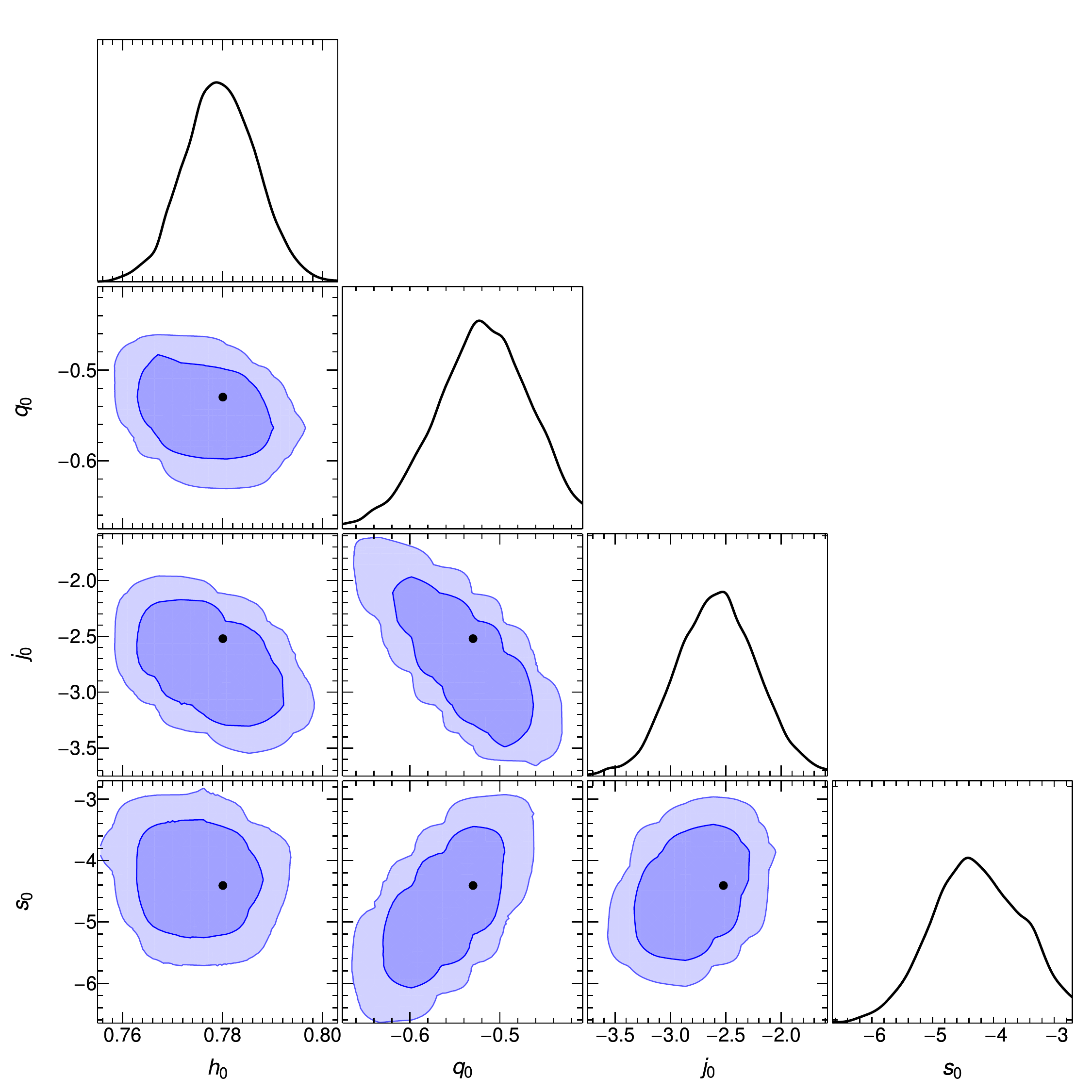}
\includegraphics[width=0.49\hsize,clip]{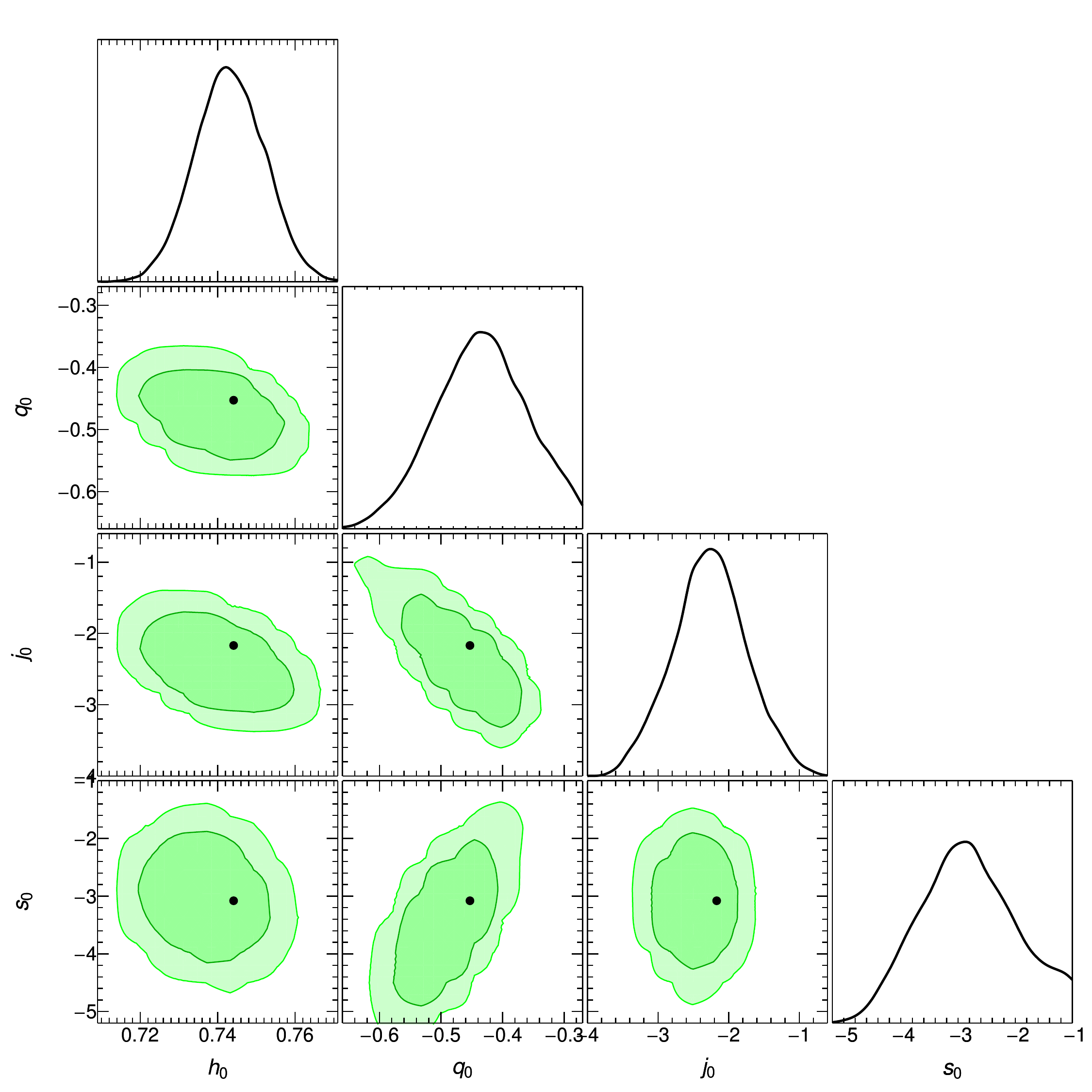}
\includegraphics[width=0.49\hsize,clip]{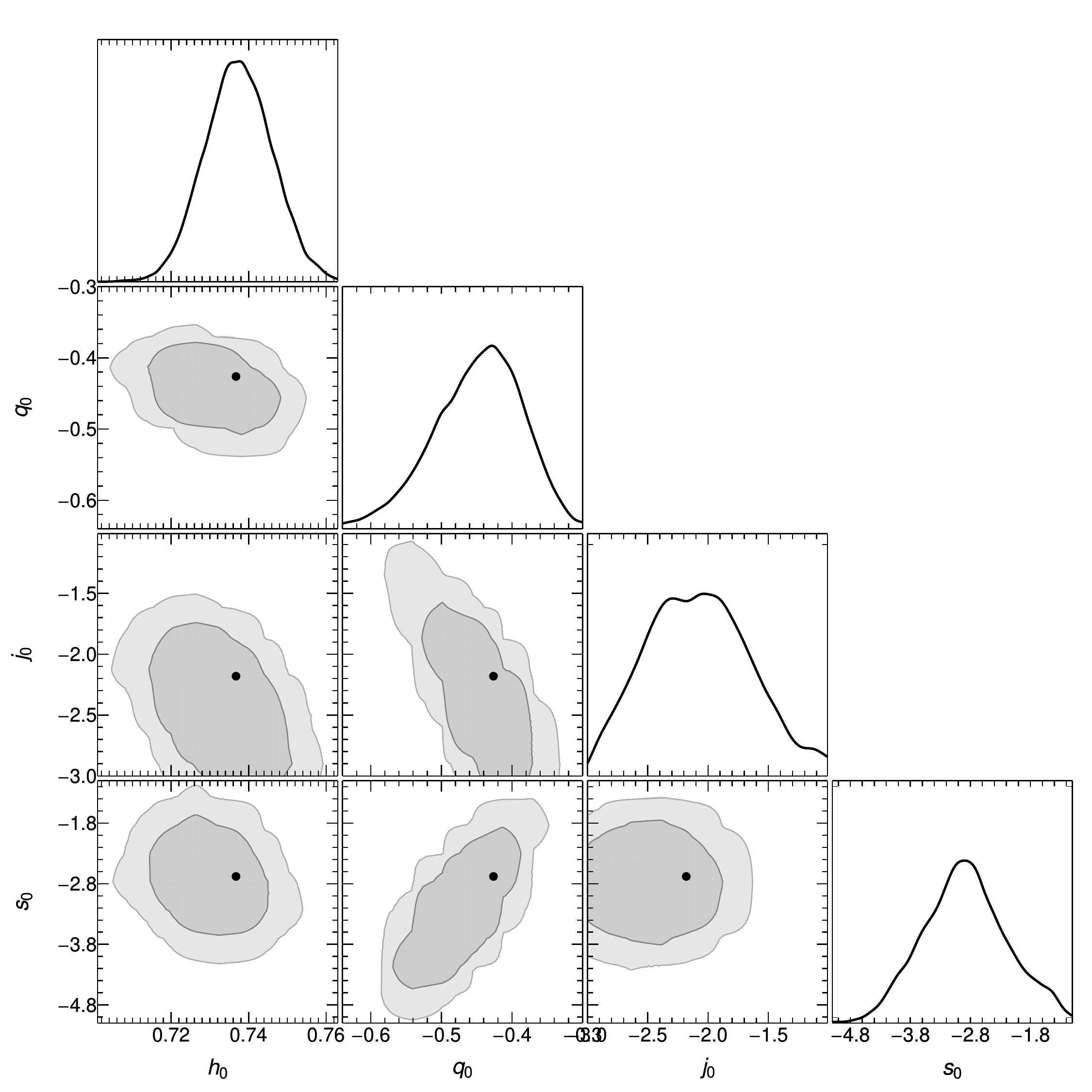}
\includegraphics[width=0.49\hsize,clip]{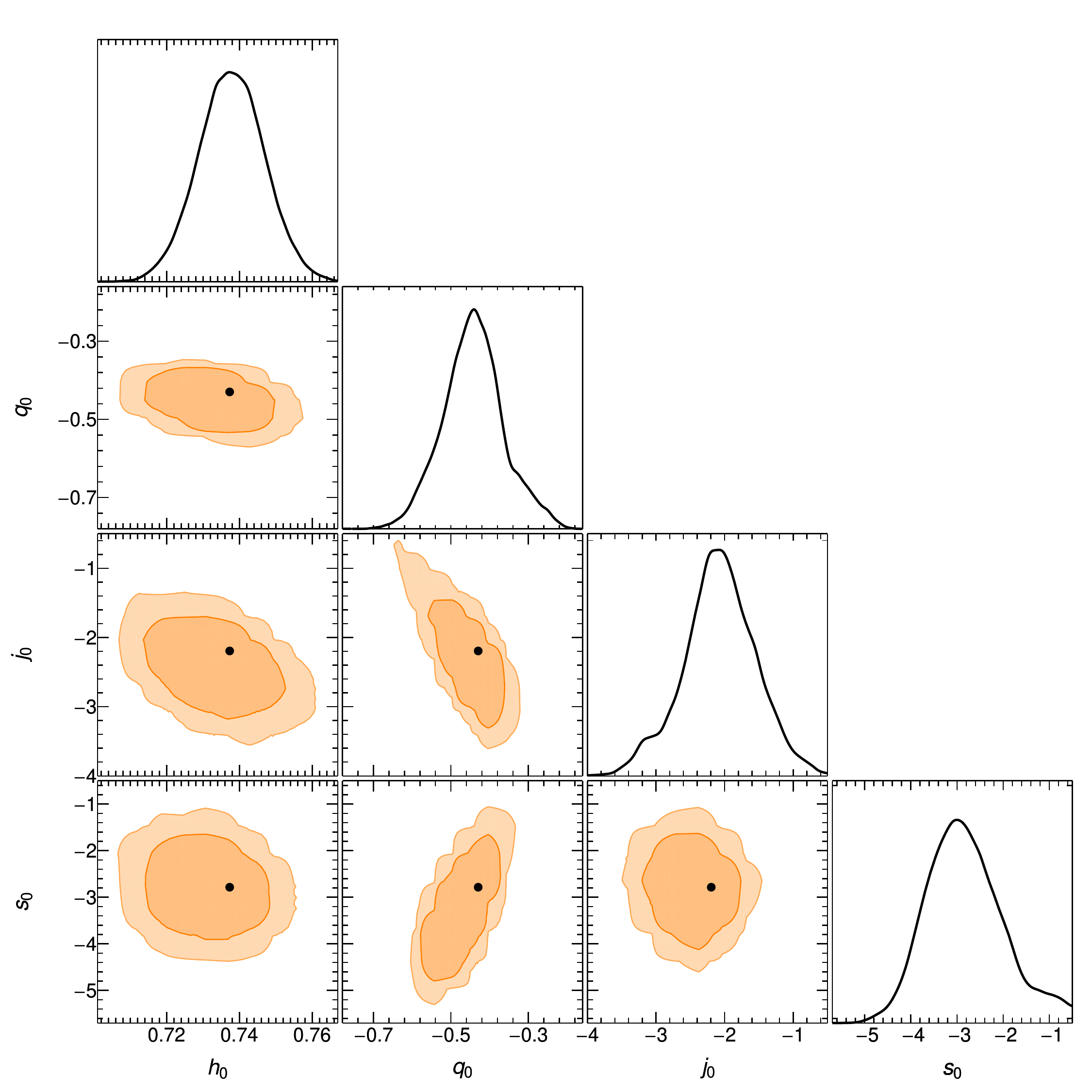}
\caption{Contours of our MCMC analyses for hierarchy 2 expansions with the auxiliary variable $y_2$. Symbols, colors, and positions of the panels are the same as in Fig.~\ref{fig:3}.}
\label{fig:7}
\end{figure*}

The adoption of the $y_2$ variable enlarges $h_0$ significantly for hierarchies 1 and 2 (see Table~\ref{tab:summary2} and Figs.~\ref{fig:6}--\ref{fig:7}, respectively). The Amati correlation indicates a quite surprising extreme value for $h_0$, meaning $h_0\sim 0.78$ for hierarchy 2. The results are, however, quite non-predictive at the level of hierarchy 1. In this case, both $q_0$ and $j_0$ are badly constrained, meaning they differ dramatically from all other fits. This signature of unpredictivity is not present as $s_0$ is introduced. In such a case, $q_0$ lies on suitable intervals, although $j_0$ and $s_0$ are very far from the $\Lambda$CDM paradigm. The main caveat with these results is that $j_0$ is negative, indicating no passage between matter and dark energy-dominated eras. Quite strangely, these results are therefore not exhaustive.

\begin{figure*}
\centering
\includegraphics[width=0.49\hsize,clip]{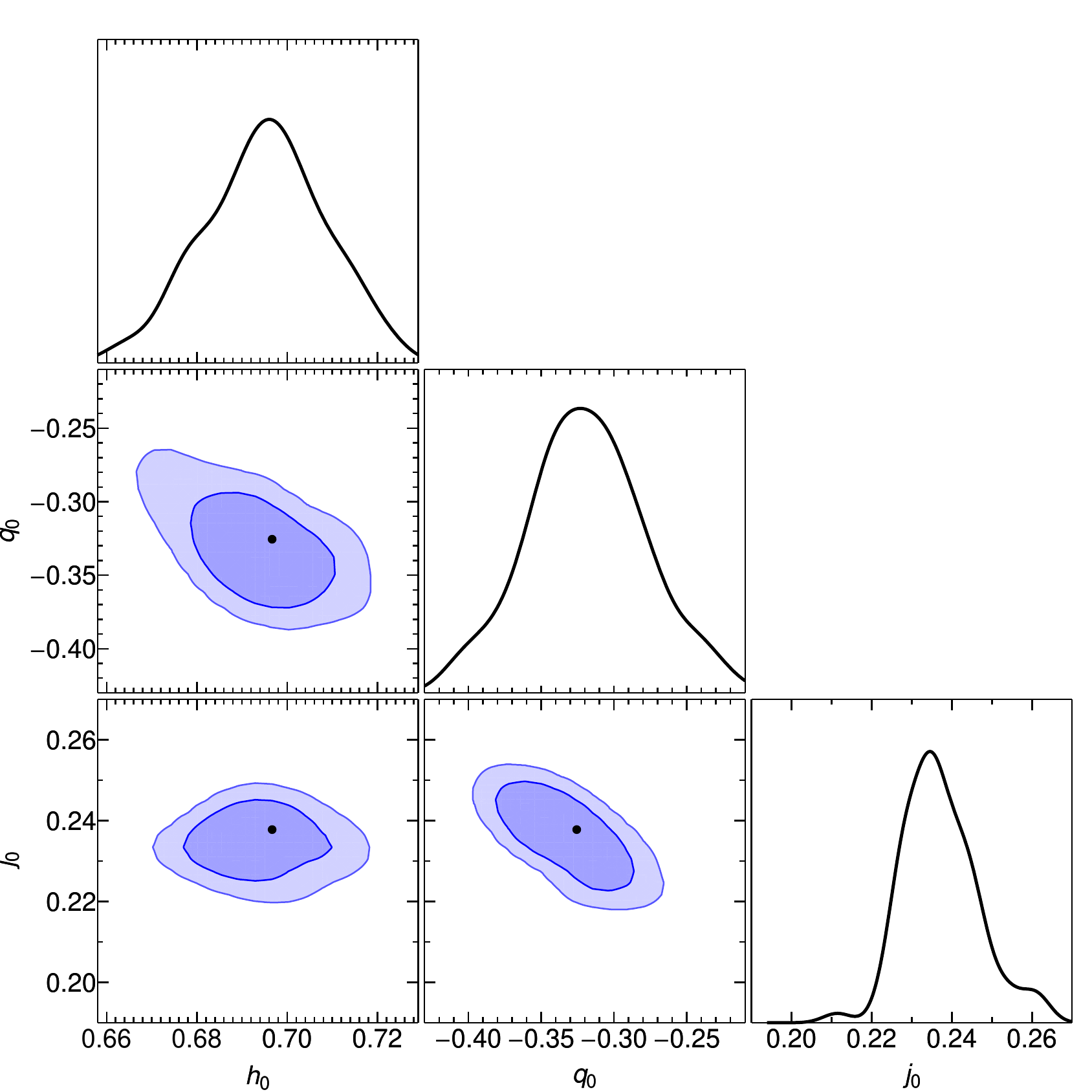}
\includegraphics[width=0.49\hsize,clip]{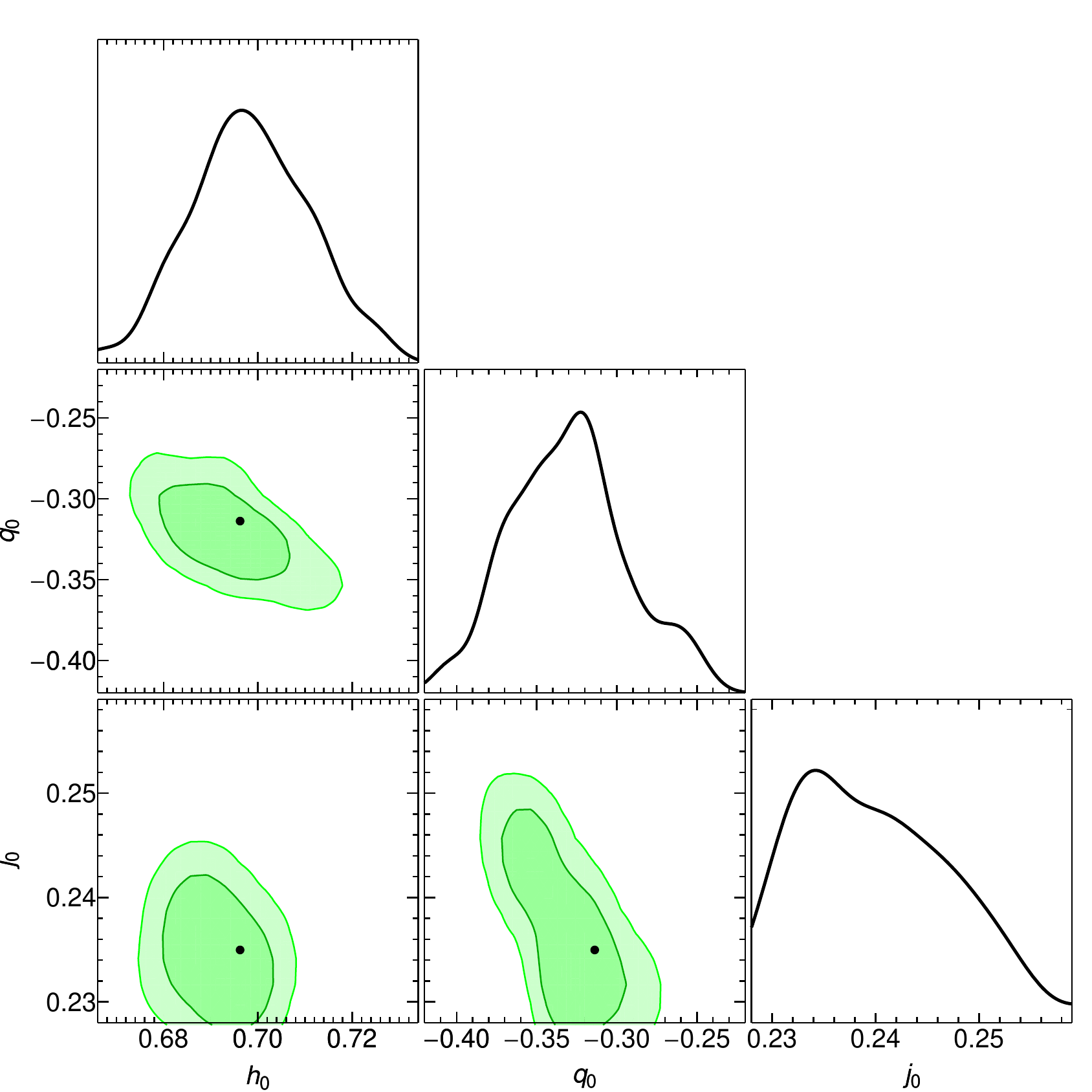}
\includegraphics[width=0.49\hsize,clip]{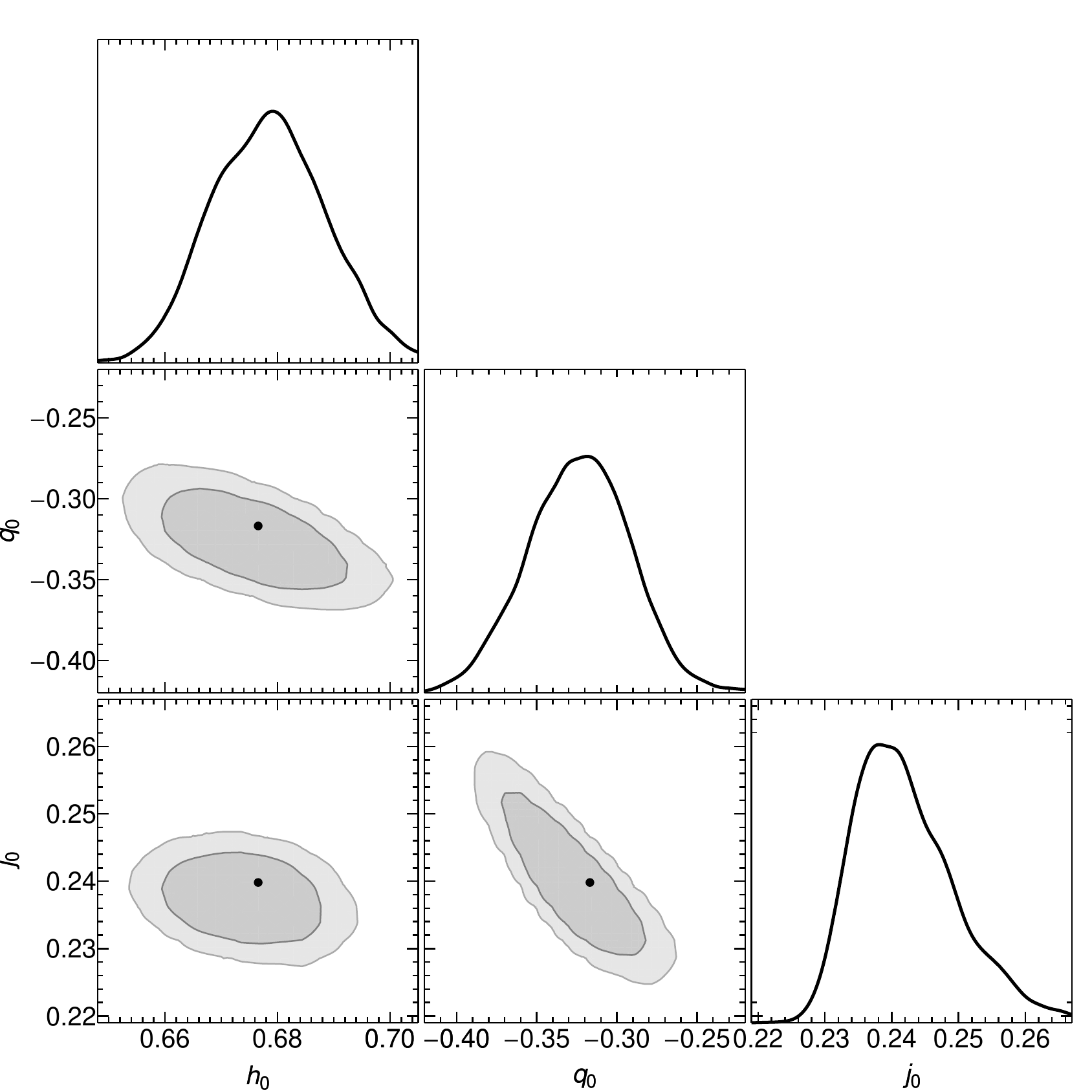}
\includegraphics[width=0.49\hsize,clip]{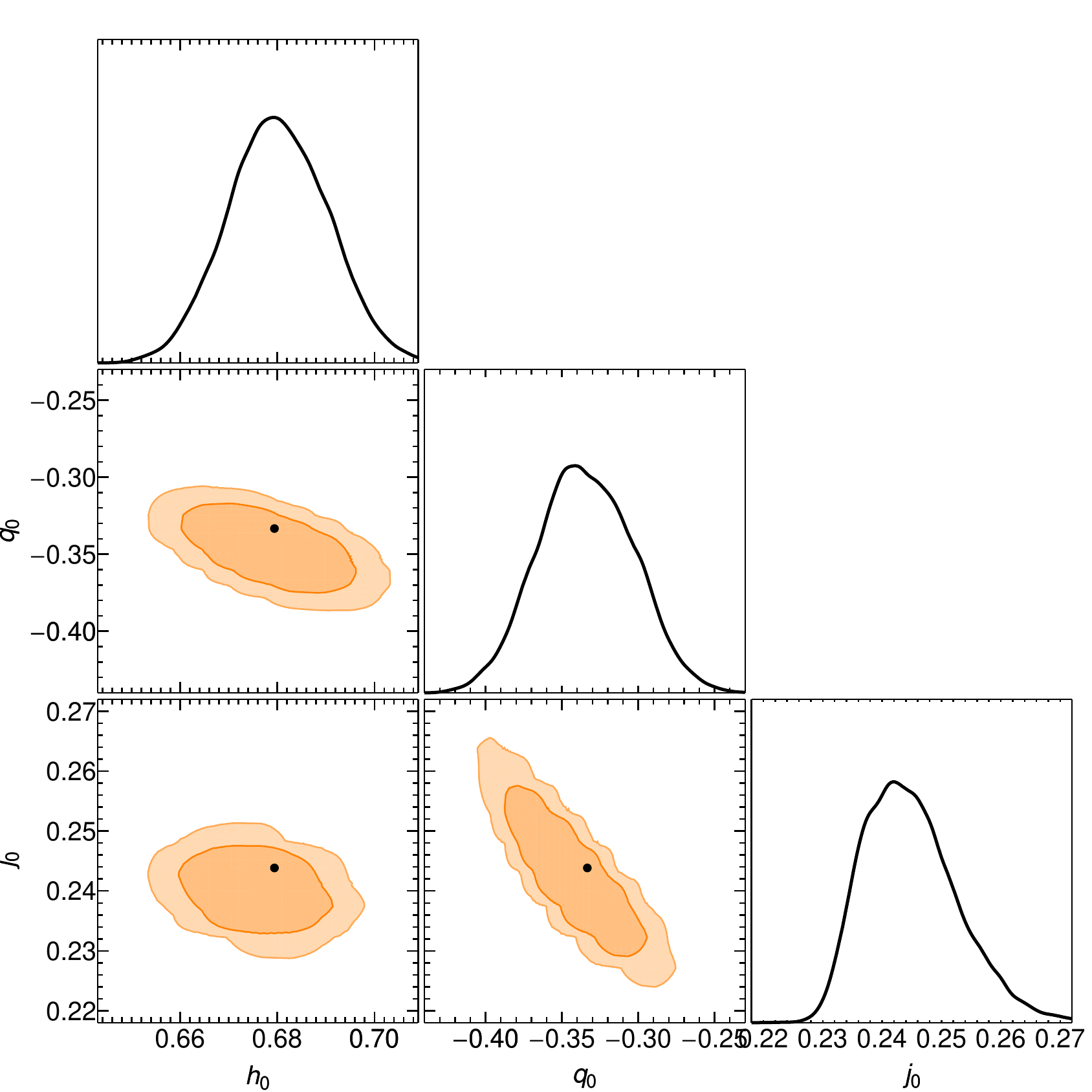}
\caption{Contours of our MCMC analyses for hierarchy 1 Pad\'e expansions. Symbols, colors, and positions of the panels are the same as in Fig.~\ref{fig:3}.}
\label{fig:5}
\end{figure*}

Pad\'e fits seem to improve the quality of hierarchy 1, as was expected by construction
(see Table~\ref{tab:summarypade} and Fig.~\ref{fig:5}).
In our fits, we find agreement with the \citet{Planck2018} expectations for $h_0$. Far from the $\Lambda$CDM approach, one notices the values of $q_0$ and $j_0$. As well as the Taylor fits, again, indications toward departures from the concordance paradigm seem to be more robust than $y_2$. Here, we notice that we could not go further to $j_0,$ since introducing $s_0$ leads to an order $\geq (3,1)$. The polynomial $P_{3,1}$ is however unconstrained at high redshift and does not perform properly as demonstrated by \citet{2020arXiv200309341C}.
\begin{table}
\centering
\setlength{\tabcolsep}{0.33em}
\renewcommand{\arraystretch}{1.5}
\begin{tabular}{l|c|c|ccc}
\hline\hline
\multirow{2}{*}{Sample}        &
\multirow{2}{*}{DoF}           &
\multirow{2}{*}{$\mathcal{A}$} & \multicolumn{3}{c}{Approximant $\chi^2$}\\
\cline{4-6}
            &
            &
            &  Taylor
            &  Function $y_2$
            &  Pad\'e $P_{2,1}$\\
\hline
\multirow{2}{*}{{\it Combo}}
            &  $1113$
            &  $1$
            &  $1116.84$
            &  $1230.71$
            &  $1113.77$\\
            &  $1112$
            &  $2$
            &  $1089.25$
            &  $1160.04$
            &\\
\hline
\multirow{2}{*}{{\it Ghirlanda}}
            &  $1080$
            &  $1$
            &  $1120.19$
            &  $1271.92$
            &  $2203.16$\\
            &  $1079$
            &  $2$
            &  $1075.01$
            &  $1184.42$
            &\\
\hline
\multirow{2}{*}{{\it Yonetoku}}
            &  $1154$
            &  $1$
            &  $1235.08$
            &  $1350.27$
            &  $1178.07$\\
            &  $1153$
            &  $2$
            &  $1147.72$
            &  $1227.25$
            &\\
            \hline
\multirow{2}{*}{{\it Amati}}
            &  $1246$
            &  $1$
            &  $2334.35$
            &  $2818.25$
            &  $2202.75$\\
            &  $1245$
            &  $2$
            &  $2174.13$
            &  $2539.98$
            &\\
\hline
\hline
\end{tabular}
\caption{$\chi^2$ values of the cosmographic fits performed over the approximants considered in this work. For each GRB correlation, we report the number of degrees of freedom (DoF) and the considered hierarchy $\mathcal{A}$. The correlations here are listed for increasing values of the ratio $\chi^2/$DoF, where the $\chi^2$ refers to the Taylor $\mathcal{A}_1$ expansion.}
\label{tab:summarychisquare}
\end{table}

Before concluding, we discuss the statistical outcomes of our analysis, as summarized in Table~\ref{tab:summarychisquare}. Looking at the $\chi^2$ values of the cosmographic fits performed over the approximants considered in this work, we notice a possible hierarchy among the correlations themselves. In particular, by sorting the correlations by the ratio $\chi^2/$DoF for Taylor $\mathcal{A}_1$ expansion (see Table~\ref{tab:summarychisquare}), we immediately see that, among the considered correlations, \textit{\emph{combo}} has the least and closest to unity ratio, while \textit{\emph{Amati}} has the largest one.
It is also worth mentioning that among the GRB correlations, the Pad\'e rational approximant $P_{2,1}$ represents a better fit with respect to the corresponding Taylor expansion $\mathcal{A}_1$ for \textit{\emph{combo}} and \textit{\emph{Yonetoku}}, while for \textit{\emph{Amati}} and \textit{\emph{Ghirlanda}} it does not. This, in principle, may reflect the nature of the above couples of correlations, meaning that and \textit{\emph{Yonetoku}} base their standardization on \textit{\emph{luminosity}} measurements, while \textit{\emph{Amati}} and \emph{\emph{\textit{\emph{Ghirlanda}} \textit{\emph{base theirs on energy}}}} values, and may in principle indicate among them the best suited observable to perform cosmological fits with GRBs.
However, though the best fit results are encouraging, from a statistical significance point of view, \textit{\emph{the Ghirlanda}} correlation has the poorest data set, meaning the number of sources fulfilling it corresponds to $\sim14\%$ of the \textit{\emph{Amati}} data set (see Table~\ref{tab:1}).  This argument in principle may apply to the \textit{\emph{\emph{\emph{combo}}}} correlation, because, after \textit{\emph{Ghirlanda}}, it has the fewest data sets among the correlations considered in this work. Future updates on these two data sets will clarify this issue and strengthen the validity of the correlations. Furthermore, another argument against the  \textit{\emph{Ghirlanda}} correlation consists of the model-dependent correction over $E_{iso}$ accounting for GRB collimated emission, which is at the core of the correlation. Thus, theoretically speaking, \textit{\emph{Ghirlanda}} should be the least bounded correlation among all cases and for both the hierarchies.
For a comparison of the results summarized in Tables~\ref{tab:summarytaylor}--\ref{tab:summarychisquare} and described in recent literature see, for example, \citet{Demianski17b}.

Concluding our discussions regarding results, it is worth mentioning that our four standardized GRB empirical correlations, despite boasting high correlation coefficients, are not fully immune to systematic errors. To better explain the role of systematics in our analyses, we must first state that all likely systematic errors entering our scenario might come from two different "domains". The first deals with the B\'ezier interpolation method that makes use of OHD data, while the second uses data directly from GRBs. Concerning the first case concerning OHD data, we can stress that these contributions have been extensively studied in the literature. In particular, in this case, data depend on stellar metallicity estimates, population synthesis models, progenitor biases, and the presence of an underlying young component in massive and passively evolving galaxies. Even though severe systematics are expected, several studies found that $H(z)$ are recovered with a $3\%$ error rate at intermediate redshifts and that systematic errors are safely kept below the statistical ones \citep{2012JCAP...08..006M,2016A&A...585A..52L,2018ApJ...868...84M}. Indeed, neglecting the systematics, the statistical errors are incorporated in GRB distance moduli through their calibrated luminosity distances and best fit correlation parameters.

Concerning GRB data, the situation becomes more complex and requires additional and more detailed considerations. Besides the intrinsic dispersions of the GRB's prompt emissions, other factors induce systematic errors. The first contribution comes from the detection from different $\gamma$-ray detectors characterized by various thresholds and spectroscopic sensitivity, which can spread relevant selection effects/biases in the correlation observables. The GRB detector trigger algorithms are based on photon counts and are number-biased against hard photons, which represent the minority in GRB spectra. This implies that, in general, a dim soft burst will be preferentially detected over a dim hard burst, creating apparent correlations where increasingly hard GRBs appear increasingly bright \citep[see, e.g.,][]{2011MNRAS.411.1843S}. Another contribution involves prompt emission again, in particular the fitting of GRB spectral energy distribution which appears to depend upon the data quality. A cut-off power-law function is an acceptable fit to GRB spectra if the detector does not have enough sensitivity to collect high-energy photons; however, the resulting $E_{\rm p}$ are always harder than the one inferred from the standard band function \citep[for details, see][]{Kaneko_2006}. A further source of bias arises from the intrinsic GRB luminosity, because GRB correlations require bright enough sources to have their actual redshifts determined, and, in fact, there is an evident lack of weak events at high redshifts \citep{2013IJMPD..2230028A}.
For hybrid correlation, for example, {\it combo}, there is a further complication in combining X-ray observables with $\gamma$-ray ones, thus mixing properties from different rest-frame energy bands. Finally, all correlations lack the physical parameter information related to their theoretical explanation. In summary, all correlations appear to be significantly affected by selection effects in trigger and spectral analyses, and redshift determinations.

In view of the above considerations, in the past literature there have been many cautions and several studies on the possible selection effects of GRB correlations, leading, however, to contrasting results \citep{Butler_2007,2008MNRAS.387..319G,2008MNRAS.391..639N,2011MNRAS.411.1843S,2009A&A...508..173A,2010PASJ...62.1495Y}.
In particular, \citet{2008PhRvD..78l3532W} noted that in four GRB correlations (including \emph{Ghirlanda} and \emph{Yonetoku,} which are considered in this work), though the statistical errors on the best fit correlation parameters are quite small, their corresponding $\chi^2$ are very large and dominated by systematic errors. The author estimated systematic errors for all correlations by requiring that $\chi^2\equiv{\rm DoF}$. Then, within $\Lambda$CDM models with different values of $\Omega_{\rm m}$, \citet{2008PhRvD..78l3532W} noted that the derived systematic errors change by less than 3\%. This result means that either the systematic uncertainties of the calibrated correlation parameters are not sensitive to cosmological parameter values, or conversely, that the inclusion (or lack thereof) of systematic errors does not significantly affect cosmological fits performed by including GRB data. In view of this result, we can quantify the influence of systematic errors on our estimates to be $\approx3$\%.
Discussions on how to go further with the concordance paradigm are the subject of the next section.

\section{Going further the concordance paradigm?}

Our current understanding of the universe lies on the assumption of the $\Lambda$CDM paradigm, which aims to describe large-scale dynamics by means of a vacuum energy cosmological constant $\Lambda$.
Its experimental estimate is, however, dramatically different from quantum field predictions. Recently, the need to revise this approach has culminated in severe discrepancies between different measures of the current Hubble rate. There is an apparent $(9\pm2.5)\%$ tension between $H_0$ measured by the \citet{Planck2018} and \citet{2018ApJ...861..126R}, respectively. Other significant evidence in favor of not yet understood new physics comes from the potential driving inflationary times, which is written in excellent approximation in terms of a Starobinsky $R+\alpha R^2$ correction of Einstein's gravity \citep{Planck2018,2020EPJP..135....1C}. The question is thus: are there new physics beyond the concordance paradigm? In our approach, we consider GRBs of which the redshift domain is far from being primordial, as for inflationary epochs, but it lies on intermediate Hubble evolution. Consequently, comparing our results and confronting them with the $\Lambda$CDM paradigm could indicate possible hints toward deviations from the standard paradigm.
Looking at our results summarized in Sec.~\ref{results}, it seems that the standard paradigm is poorly constrained and fails to be predictive when GRBs are involved. In particular, while the auxiliary variables are far from being suitable benchmarks of our conclusions, both Taylor and Pad\'e approximations show  that $q_0$ and $s_0$ are roughly comparable to the $\Lambda$CDM model, whereas $j_0$ is not. The issue over $H_0$ is unsolved since many measures seem to agree with the \citet{Planck2018} expectations, but many others with those of \citet{2018ApJ...861..126R}. We conclude that, besides the $H_0$ tension, the concordance paradigm is only partially recovered if one assumes the above results are not jeopardized by high systematics. In this case, we discussed above that it might not contribute significantly to the overall expectations. However, even if our indications seem to be in favor of a model practically comparable to the $\Lambda$CDM approach, but with varying $j$, more accurate studies are needed to check whether systematics really affect the results or not. Even though this represents a suitable possibility, another aspect to be considered could be to take a spatial curvature different from zero. This scenario arises because $s$ can also vary according to time differently to the way it does so according to the $\Lambda$CDM framework, and our bounds do not, \emph{a priori,} exclude the occurance of slight departures.
This is why the speculation of non-negligible spatial curvature would need to be intensively investigated to exclude or confirm whether GRBs with different calibrated correlations are in favor of a spatially curved universe \citep{2020ApJ...888...99W}.


\section{Final outlooks and perspectives}

In this paper, we revised the Universe's dynamics by using high-redshift data from GRBs. To do so, we constrained cosmographic parameters by means of model-independent techniques through four different GRB calibrations. To overcome the issue of circularity, we proposed a recent new method, based on Bezi\'er polynomials. These  have been employed to set correlations to heal the circularity problem, using OHD data catalogs. The main advantage of our method is that it frames our analysis in a model-independent way, healing, \emph{de facto,} the circularity problem without postulating the model \emph{a priori}. With the model-independently calibrated \emph{Amati}, \emph{Ghirlanda}, \emph{Yonetoku}, and \emph{combo} correlations, several MCMC analyses were  computed to obtain$1$--$\sigma$ and $2$--$\sigma$ confidence levels and to test the standard cosmological model.
To do so, we considered the most recent approaches to cosmography, comparing, for example, Taylor expansions with $z$ and $y_2$ series, and Pad\'e polynomials. We derived limits over the Universe's expansion history, fixing constraints over $h_0, q_0, j_0,$ and $s_0$ within two hierarchies, $\mathcal{A}_1$ up to $j_0$ and $\mathcal{A}_2$ up to $s_0$. We used GRB data from the aforementioned four calibrated GRB correlations, together with SNeIa and BAO catalogs. Reasonable results have been found for $\mathcal{A}_1$ and $\mathcal{A}_2$ hierarchies through several MCMC fits, demonstrating possible matching with the standard paradigm. Moreover, we only partially alleviated the tension on local $H_0$ measurements where the $\mathcal{A}_2$ hierarchy is considered. Quite surprisingly, our findings showed that the $\Lambda$CDM model is not fully confirmed using GRBs. Although this can be an indication that more refined analyses are required where GRBs are involved, simple indications seem to be against a genuine cosmological constant and are interpreted throughout the text, either with a barotropic dark energy contribution, or with the need for nonzero spatial curvature.
Nevertheless, at this stage, our findings are in line with recent claims regarding tensions with the $\Lambda$CDM model \citep{2019arXiv191101681Y,2019A&A...628L...4L,2019NatAs...3..272R}. In this respect, we also developed comments on previous works concerning the role of systematic errors. In particular, we studied how systematics can affect our results in view of the cosmographic bounds. We found that no significant deviations are expected in agreement with previous works discussed on the same topic.

Future perspectives will shed light on the role of spatial curvature. We will resolve the experimental fits in which spatial curvature is not fixed to zero at the very beginning. Moreover, we intend to adapt our approach, based on B\'ezier polynomials, directly to the case of dark energy paradigms. We endeavour to investigate this possibility in order to verify, using GRB data, whether or not our obtained deviations are somehow reproducible in the context of dark energy scenarios.

\begin{acknowledgements}
OL and MM acknowledge the INFN support, and in particular the Frascati National Laboratories, for Iniziative Specifiche MOONLIGHT2. OL acknowledges the support provided by the Ministry of Education and Science of the Republic of Kazakhstan, Program IRN: BR05236494 for support. OL and MM are thankful to Hirotaka Ito for sharing the updated GRB data of the \emph{Yonetoku} correlation. The work is dedicated to the lovely memory of our colleague, Diego Leonardo Caceres Uribe, excellent scientist and good friend.
\end{acknowledgements}

\bibliographystyle{aa}
\bibliography{biblio,bibliomiaaa}

\clearpage

\appendix

\end{document}